\documentclass[11pt]{article}
\usepackage{jheppubmod}
\pdfoutput=1
\usepackage{mathtools}
\mathtoolsset{showonlyrefs}
\usepackage{psfrag}
\usepackage{array}
\usepackage{amssymb}
\usepackage{amsmath}
\usepackage{amsthm}
\usepackage{centernot}
\usepackage{graphicx}
\usepackage{epstopdf}
	
\usepackage{epsfig}
\usepackage{wrapfig}
\usepackage[punctsep]{collref}
\collectsep[]{;}
\def\bconst{\kappa}
\def\op{O}
\def\hilbe{H_E}
\def\pl{\text{Pl}}

\def\al{A}

\def\pb[#1,#2]{\{#1, #2\}}
\def\deb[#1,#2]{[#1,#2]_{\text{D.B.}}}

\def\cn{{\mathcal N}}
\def\tr{{\rm Tr}}

\def\Or[#1]{{\text{O}}\left({#1}\right)}
\def\dotl[#1,#2]{\left\langle #1,\, #2 \right\rangle}
\def\dotlb[#1,#2]{\left\langle #1,\, #2 \right\rangle}
\def\dotlm[#1,#2]{\left[ #1,\, #2 \right]}
\def\dotp[#1,#2]{(\vect{#1} \cdot\vect{#2})}
\def\aff[#1,#2]{\hat{#1}(#2)}

\def\n4sym{{\cal N}=4 SYM}
\def\>{\rangle}
\def\<{\langle}
\def\weight[#1,#2,#3]{\{(#1),#2,#3\}}
\def\ads[#1]{$\text{AdS}_{#1}$}

\newcommand{\Airy}{\operatorname{Ai}}

\newcommand{\lpl}{\ell_{\text{pl}}}
\newcommand{\ls}{\ell_{\text{str}}}

\newcommand{\be}{\begin{equation}}
\newcommand{\ee}{\end{equation}}
\newcommand{\ba}{\begin{align}}
\newcommand{\ea}{\end{align}}
\newcommand{\bs}{\begin{split}}
\def\sess\end{split}
\newcommand{\vect}[1]{{\boldsymbol{#1}}}

\def\opbulk{\hat{O}}
\def\opbulkbh{O^{\text{bh}}}
\def\opbulkfuzz{O^{\text{fuzz}}}
\def\fm{f}
\def\none{N_1}
\def\nfive{N_5}
\def\smfunc{{\mathfrak{F}}}

\def\expgenfn[#1,#2]{{\cal G}(#1, #2)}
\def\expgenfnbeta[#1,#2]{{\cal G}_{\beta}(#1, #2)}
\def\fluctgenfnbeta[#1,#2,#3,#4]{{\cal V}_{\beta}(#1,#2,#3,#4)}
\newcommand{\exponfn}{{\cal F}}
\newcommand{\exponfnpr}{\widetilde{\cal F}}
\newcommand{\rstretch}{{R_{\scriptscriptstyle{\rm{stretch}}}}}
\newcommand{\rstretchsq}{{R^2_{\scriptscriptstyle{\rm{stretch}}}}}
\newcommand{\vcom}{V_{\scriptscriptstyle{\rm com}}}
\def\normconst{\cal N}
\def\thermexp[#1]{{\langle #1 \rangle_{\beta}}}
\def\typexp[#1]{{\langle \Psi| #1 | \Psi \rangle}}
\def\sigmaens{\sigma_{\text{ens}}}

\newtheorem{result}{Result}
\newtheorem{expectation}{Expectation}
\def\tauconst{\tau} 
\newcommand{\quantpar}{\mathsf{q}}
\newcommand{\devpar}{\mathsf{d}}
\def\wight{G}
\def\comm{J}
\def\casympt{C}
\def\momry{\gamma}
\def\freqry{\omega}
\def\psifunc{\psi}
\def\normalpha{\alpha}
\def\chip{\tilde{\chi}}
\def\alphap{\tilde{\alpha}}
\def\gp{\tilde{g}}
\def\tp{\tilde{t}}
\def\sp{\tilde{s}}
\def\wg{W}
\def\ximid{\xi_{\text{m}}}

\title{A Critique of the Fuzzball Program}
\author{Suvrat Raju and Pushkal Shrivastava}
\emailAdd{suvrat@icts.res.in}
\emailAdd{pushkal.shrivastava@icts.res.in}
\affiliation{International Centre for Theoretical Sciences, Tata Institute of Fundamental Research, Shivakote, Bengaluru 560089, India.}
\abstract{We explore the viability of fuzzballs as candidate microstate geometries for the black hole, and their possible role in resolutions of the information paradox. We argue that if fuzzballs provide a description of black-hole microstates, then  the typical fuzzball geometry can only differ significantly from the conventional black-hole geometry at a Planck-scale-distance from the horizon. However, precisely in this region, quantum fluctuations in the fuzzball geometry become large and the fuzzball geometry becomes unreliable. We verify these expectations through a detailed calculation of quantum expectation values and quantum fluctuations in the two-charge fuzzball geometries. We then examine some of  the solutions discovered in arXiv:1607.03908. We show, based on a  calculation of a probe two-point function in this background, that these solutions, and others in their class, violate robust expectations about the gap in energies between successive energy eigenstates,  and differ too much from  the conventional black hole to represent viable microstates.  We conclude that while fuzzballs are interesting star-like solutions in string theory, they do not appear to be relevant for resolving the information paradox, and cannot be used to make valid inferences about black-hole interiors.}

\begin{document}
\maketitle
\section{Introduction}
In higher dimensional supergravities, it is sometimes possible to find horizonless solutions, called fuzzballs,  with the same charges as a black hole. Over the past few years, considerable effort has been devoted to discovering and analyzing new classes of fuzzball solutions. Such solutions can be found both in asymptotically flat space and in asymptotically anti-de Sitter space.

The fuzzball program (see \cite{Mathur:2005zp,Bena:2007kg,Bena:2013dka} and references there)  is the bold idea that such geometries can be used to parameterize the set of microstates in quantum gravity that correspond to a black hole. The fuzzball program suggests that the black hole should be viewed only as some kind of ``average'' geometry, with individual microstates specified by distinct horizonless geometries. This has two immediate implications: first, that the entropy of the black hole can be recovered by quantizing the moduli-space of fuzzball solutions and second, that the black hole truly has no interior.

The fuzzball program has not been carried through to completion in any setting corresponding to a macroscopic black hole. The moduli-space of fuzzballs has only been quantized \cite{Rychkov:2005ji} for fuzzballs that are dual to ground states of the D1-D5 system \cite{Lunin:2001fv,Lunin:2002bj}, which do not correspond to a black hole with a  macroscopic horizon. Nevertheless, given the considerable effort that continues to be directed towards understanding fuzzballs, we believe it is pertinent to address the following question: {\em is it consistent with the principles of statistical-mechanics to expect that black-hole microstates can be represented by distinct geometries, which can be analyzed by studying classical solutions}?

In essence our analysis is very simple. First, we point out that quantum-mechanical microstates in a system with $e^{S}$ states are almost all indistinguishable from one-another: when microstates are probed with any reasonable observable,  the differences between distinct typical states are of size $e^{-{S \over 2}}$.  For black holes, where $S$ scales as an inverse power of the Planck-scale,  we cannot possibly expect to represent the exponentially small difference between two typical microstates using geometry. Therefore, we argue that microstates that occupy almost all of the volume of Hilbert space must be represented by a single universal geometry, and we further argue that this  must be the conventional black-hole geometry.

Then we consider the weaker possibility that while fuzzballs might not provide a useful representation of typical states, they might still provide a basis that spans all black-hole microstates. In this context, we prove some simple bounds on how atypical basis elements can be.   We use these bounds to show that if fuzzballs are to form even only a basis, then in most of space, typical fuzzball metrics must equal the conventional black-hole metric  up to terms that are suppressed by ${1 \over \sqrt{S}}$. So in most regions of space such fuzzballs are effectively indistinguishable from the black hole.

We further argue that if typical fuzzballs are to deviate significantly from the conventional black hole, they can only do so once we are within Planck length of the horizon. We show that this deviation {\em cannot occur at a larger scale like the string-scale.} Therefore, if fuzzballs are to be microstates, then typical fuzzballs must be represented by metrics with explicit Planck-scale structures. We argue that such metrics cannot be analyzed classically --- either in supergravity or in classical string theory --- since quantum fluctuations in such metrics are of the same order as the classical Planck-scale structures. Therefore, in the region where fuzzballs may have displayed interesting deviations from the black hole, they are unreliable. 

In fact, several of the explicit fuzzball solutions that have been found have structures at macroscopic scales --- much larger than then Planck scale. (In the terminology of \cite{Bena:2013dka}, such solutions are called ``microstate geometries''.)  We argue that such fuzzballs are irrelevant to the discussion of black-hole microstates since they do not satisfy our bounds on how close elements of a basis have to be to the ensemble average.

The arguments above are explained in greater detail in section \ref{secstatprelim}. It is sometimes argued that fuzzballs are required to resolve the information paradox \cite{Mathur:2008nj,Mathur:2009hf,Mathur:2012np}, and we explain, in section \ref{subsecindirectarg}, why we believe that this argument is incorrect.

We believe that our arguments are robust but, in order to check these arguments, we performed several detailed calculations with explicit examples of fuzzball solutions. These calculations, which are described in sections \ref{sectwocharge} and \ref{secmulticharge} take up the bulk of this paper; the reader who is not persuaded by our abstract arguments in section \ref{secstatprelim} should consult these concrete examples.

In section \ref{sectwocharge} we analyze the Lunin-Mathur geometries corresponding to ground states in the D1-D5 system \cite{Lunin:2002bj,Lunin:2001jy} that were quantized by Rychkov \cite{Rychkov:2005ji}. In this quantum-statistical system, we are able to verify our general expectations. We can compute both the quantum-mechanical expectation of components of the metric, and also fluctuations in these quantities. We show that the geometry corresponding to a {\em typical} microstate differs from the conventional solution only at the Planck scale. 
Moreover, in the region where the geometry differs in an interesting manner, the fluctuations in physical quantities become the same size as their classical expectation values and the geometry becomes unreliable, precisely in line with our general expectations.

In section \ref{secmulticharge} we analyze the recently discovered class of asymptotically AdS solutions that correspond to 1/4-BPS states in the D1-D5 system \cite{Bena:2016ypk}. Such 1/4-BPS states are described by a black hole with finite horizon area \cite{Cvetic:1998xh} but the geometries of \cite{Bena:2016ypk} differ macroscopically from the black-hole geometry. We show that these differences can be easily detected through simple asymptotic boundary observables. 

More specifically, in these geometries, we compute the two-point Wightman function and commutator of a marginal boundary scalar operator. We show that the support of these  functions in frequency space is concentrated on a set of discrete well-separated frequencies, in contradiction with what one would expect for a black-hole microstate --- where we expect the support to be effectively continuous in frequency space. Next, we show that the falloff of these functions for large spatial momenta in the fuzzball geometry fails to saturate a bound that is saturated by the black-hole geometry. These calculations provide strong evidence that the geometries of \cite{Bena:2016ypk}  are not viable as typical microstates.

While our calculations in section \ref{secmulticharge} are specific to a class of solutions, we believe that our conclusions are far more general. If a geometry has macroscopic features, these macroscopic features can be detected by appropriate asymptotic correlation functions, and they lead to violations on the bounds of how atypical a basis for black-hole microstates can be.

\begin{figure}
\includegraphics[width=\textwidth]{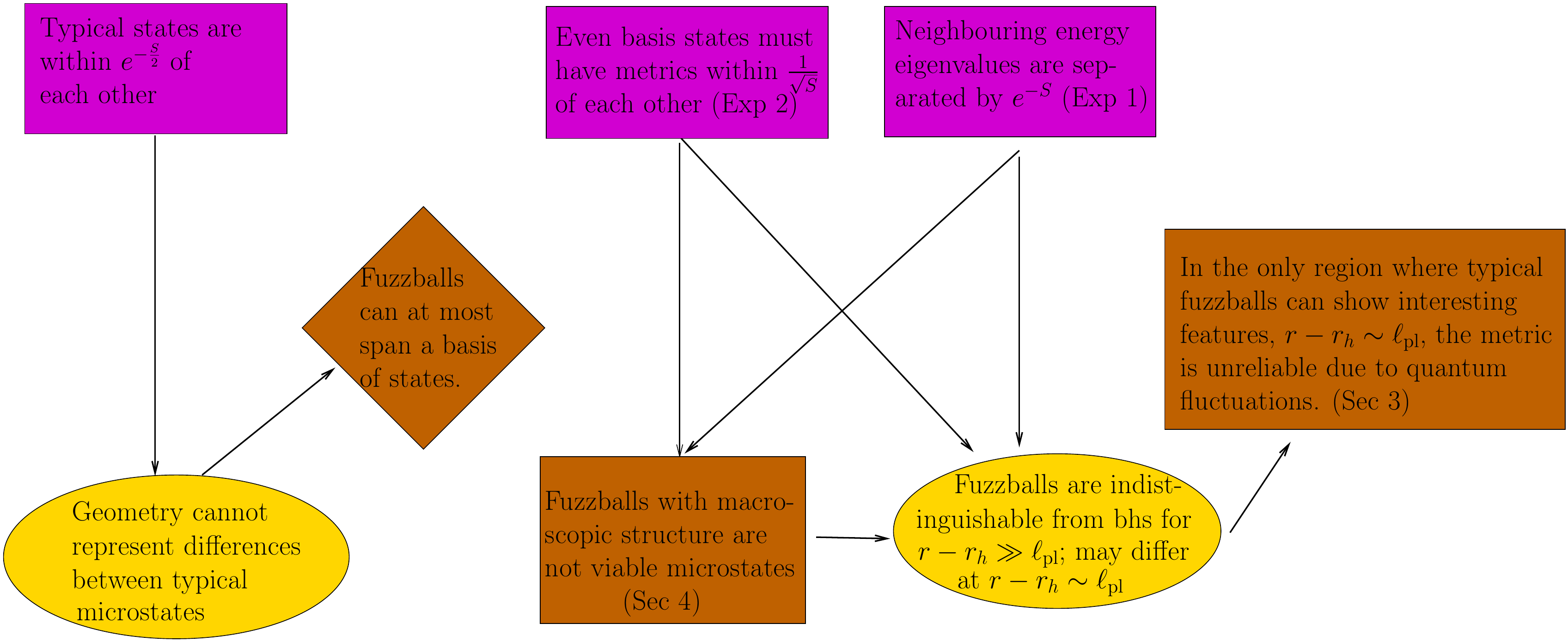}
\caption{\em Logical flowchart of the paper. Statistical mechanics results are in magenta rectangles. The physical expectation that they rely on is referred to in brackets.  Implications for the fuzzball are in orange ovals. Major conclusions are in brown boxes. Sections where a conclusion is verified are given in brackets. \label{logicchart}}
\end{figure}

Figure \ref{logicchart} outlines the flow of logic in this paper and explains how our calculations in sections 3 and 4 fit into this flow.
\section{A statistical-mechanics evaluation of the fuzzball program  \label{secstatprelim}}
In this section, we review some simple results from quantum statistical mechanics, and explain their implications for the fuzzball program. These results will motivate our calculations in section \ref{sectwocharge} and \ref{secmulticharge}. We have organized this section into three subsections: in subsection \ref{subsecstatresults} we review some results from statistical mechanics; in subsection \ref{subsecimplicationfuzz} we explain the relevance of these results for the fuzzball program; in subsection \ref{subsecindirectarg} we discuss the ``Hawking theorem'' described in \cite{Mathur:2009hf} which is sometimes used to indirectly infer properties of fuzzballs.  

Some readers may be concerned that our arguments in this section are too abstract. We urge these readers to read this section in conjunction with section \ref{sectwocharge} and section \ref{secmulticharge} where we have performed a number of calculations that support our deductions in specific examples.

\subsection{Some results from statistical mechanics \label{subsecstatresults}}
We now discuss some results that characterize (a) typical states in high-dimensional quantum statistical systems (b) the extent to which elements of a complete basis can differ from one another and (c) the gap between neighbouring energy eigenstates.
 
\begin{result}
\label{microtyp}
Consider any subspace $\hilbe$ of a Hilbert space. Let $\text{dim}(\hilbe) =  e^{S}$ and let $\mu_{\psi}$ be the Haar measure on $\hilbe$ in the neighbourhood of a state $|\psi \rangle$.  Then typical pure states in $\hilbe$ are exponentially close to the maximally mixed state on $\hilbe$ in the sense that for any Hermitian operator $\al$, we have 
\be
\label{hilbtypical}
\langle \al \rangle \equiv \int \langle \Psi | \al | \Psi \rangle d \mu_{\psi} = \tr(\rho \al),
\ee
where the density matrix, $\rho = e^{-S} P$, and $P$ is the projector onto $H_{E}$.  Moreover, deviations from this mean value are  exponentially suppressed
\be
\int \left(\langle \Psi | \al | \Psi \rangle - \langle \al \rangle \right)^2 d \mu_{\psi} \leq {\sigmaens^2 \over e^{S} + 1},
\ee
where $\sigmaens^2 \equiv \tr(\rho \al^2 ) - \big[\tr(\rho \al)\big]^2$.
\end{result} 

To our knowledge this result was first described in \cite{lloyd1988black}. To prove this result we choose some basis for the subspace, and we label its elements by $|\fm_1\rangle, |\fm_2 \rangle \ldots |\fm_{e^{S}} \rangle$. Then an arbitrary state in this subspace can be written as $|\Psi \rangle = \sum_{i} a_i |\fm_i \rangle$. The Haar measure is given by
\be
\label{haarhilb}
d \mu_{\Psi} = {1 \over V} \delta(\sum_{i=1}^{e^S} |a_i|^2 - 1) \prod_{j=1}^{e^S} d a_j,
\ee
where $V$ is a normalization-constant which can be set by demanding that $\int d \mu_{\Psi} = 1$, which leads to
$V^{-1} =  {\pi^{e^S} \over \Gamma(e^S)}$.  We emphasize that the measure \eqref{haarhilb} is {\em independent} of the choice of basis.

Now consider an arbitrary Hermitian operator, $\al$ and denote its matrix elements in the basis above by $\al_{i j} = \langle \fm_j | \al | \fm_i \rangle$.  Then
\be
\int \langle \Psi | \al | \Psi \rangle d \mu_{\psi} = \int d \mu_{\psi} \Big[\sum_{i = 1}^{e^S} |a_i|^2 \al_{i i}  + \sum_{i \neq j} a_i a_j^* \al_{i j} \Big] 
= {1 \over e^{S}} \sum \al_{i i} = \tr(\rho \al),
\ee
where we have used the fact that $\int d \mu_{\psi} a_i a_j^* = {1 \over e^{S}} \delta_{i j}$.

A simple computation yields the variance in the second part of the result.
\be
\begin{split}
&\int \Big[\langle \Psi | \al | \Psi \rangle  - \tr(\rho \al) \Big]^2 d \mu_{\psi} = \int \Big[\sum_{i,j} \al_{i j} a_i a_j^* - \sum_{i} \al_{i i} |a_i|^2 \Big]^2 \\
&= \int \Big[\sum_{i \neq j, l \neq m}  \al_{i j} \al_{l m} a_i a_j^* a_l a_m^* \Big] d \mu_{\psi} 
= \int \Big[\sum_{i \neq j} |\al_{i j}|^2 |a_i|^2 |a_j|^2\Big] d \mu_{\psi} \\
&=  {1 \over e^{S} (e^{S} + 1)} \sum_{i \neq j} |\al_{i j}|^2 
\leq {1 \over e^{S} + 1} \sigmaens^2.
\end{split}
\ee
Here, in the second line we used the fact that unless $i = m$ and $j = l$, the summand vanishes upon integration. In the third line, we used the fact that $\int d \mu_{\psi} |a_i|^2 |a_j|^2  = {1 \over e^{S} (e^{S} + 1)}$ for $i \neq j$. A small subtlety in the final step is that 
\be
{1 \over e^{S}} \sum_{i \neq j} |\al_{i j}|^2  = \tr(\rho (P \al P)^2) - \tr(\rho \al)^2 \leq \sigmaens^2.
\ee
This difference arises because $\al$ might have matrix elements that link states in $\hilbe$ to states outside $\hilbe$. 
\begin{figure}
\begin{center}
\includegraphics[width=0.3\textwidth]{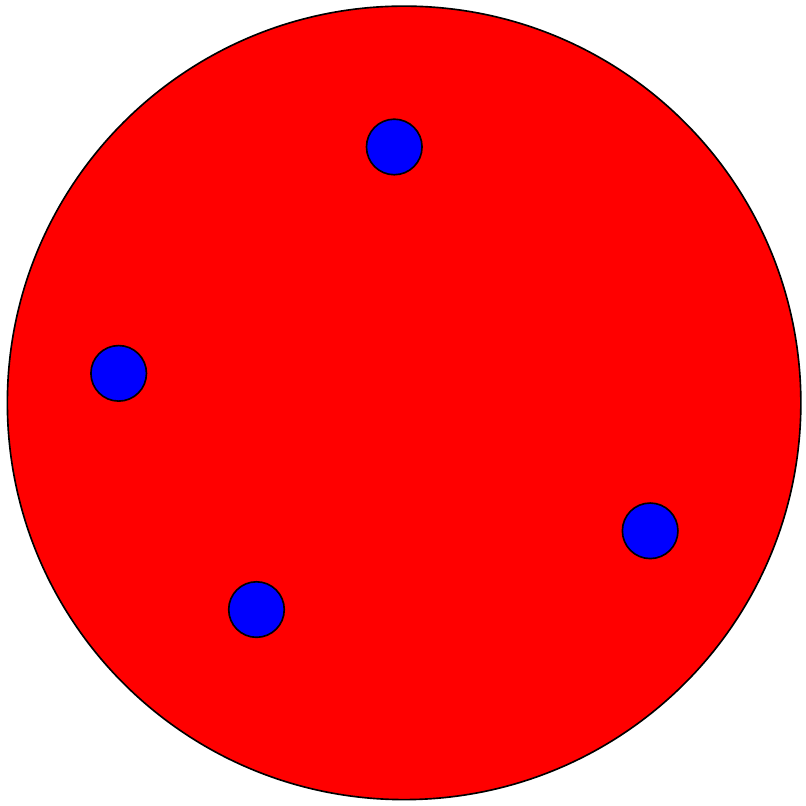}
\caption{\em The subspace, $\hilbe$, is a compact manifold. Most pure states in the space are very close to the maximally mixed state. An exponentially small volume of states (displayed in blue) can be atypical.\label{fighilbtypical}}
\end{center}
\end{figure}

This result should be interpreted as follows: {\em ``On almost all of the volume of the subspace, the expectation value of the operator differs from the typical expectation value by an exponentially small amount. The expectation value may differ significantly from the typical expectation value only an exponentially small region of the subspace.''} The reader may consult Figure \ref{fighilbtypical} for intuition. \allowbreak

This result tells us that {\em typical} microstates of $\hilbe$ are described by a universal set of correlators. By itself, this does not disallow the possibility of an atypical basis of states for $\hilbe$.
This is because the basis vectors themselves occupy only zero volume in the Hilbert space. We now bound the atypicality of a basis in some cases of interest.

\begin{result} (Limit on atypicality of a basis)
\label{thatypical}
Assume that the ratio  ${\sigmaens \over \langle \al \rangle}$ vanishes  as  ${1 \over S^{\alpha}}$  for large $S$ and some positive number $\alpha$. Given any basis, $|\fm_1 \rangle \ldots |\fm_{e^S} \rangle$, for $\hilbe$, let $|\fm_{\alpha_1} \rangle \ldots |\fm_{\alpha_M} \rangle$ be those of its elements where $\Big|{\langle \fm_{\alpha_j} | \al | \fm_{\alpha_j} \rangle - \langle \al \rangle \over \langle \al \rangle}\Big| \geq \Or[{1 \over S^{\beta}}]$ in the thermodynamic limit, with $\beta < \alpha$.  Then ${M \over e^{S}}$ vanishes at least as fast as $\Or[{1 \over S^{2 (\alpha - \beta)}}]$. 
\end{result}
This result follows from simple inequalities. 
\be
\begin{split}
\sigmaens^2 
&= {1 \over e^{S}} \sum_i \langle \fm_i | \al^2 | \fm_i \rangle   - \langle \al \rangle^2   \\
&= {1 \over e^{S}} \sum_i \left(\langle \fm_i | \al^2 | \fm_i \rangle  - \langle \fm_i | \al | \fm_i \rangle^2 \right) + {1 \over e^{S}} \sum_i \left(\langle \fm_i| \al | \fm_i \rangle - \langle \al \rangle \right)^2 \\
& \geq {1 \over e^{S}} \sum_{j = 1}^M \left( \langle \fm_{\alpha_j} | \al | \fm_{\alpha_j} \rangle - \langle \al \rangle \right)^2 \geq {M \kappa^2 \langle \al \rangle^2 \over e^{S}},
\end{split}
\ee
where $\kappa = \text{inf}_{j} \Big|{\langle \fm_{\alpha_j} | \al |\fm_{\alpha_j} \rangle| - \langle \al \rangle \over \langle \al \rangle} \Big|$. By assumption $\kappa = \Or[{1 \over S^{\beta}}]$ in the thermodynamic limit, and since ${\sigmaens^2 \over \langle \al \rangle^2}$ vanishes like ${1 \over S^{2 \alpha}}$,  therefore ${M \over e^{S}}$ must vanish like ${1 \over S^{2 (\alpha - \beta)}}$. 

The result above is very simple, but it is relevant for those observables that take on a finite {\em classical expectation value}. These are the observables where ${\sigmaens \over \langle \al \rangle}$ vanishes as $S \rightarrow \infty$. For such observables, the result states that one cannot construct a basis whose elements are all individually very different, and only average out to give some mean.

So far our results have been kinematical. We now state a dynamical  {\em expectation} about the spectrum, which should be true in almost all interacting systems. Let $S$ be the entropy at energy $E$. (We deliberately use the same notation as above since $e^{S} = \dim(\hilbe)$ if $\hilbe$ is taken to the subspace corresponding to the microcanonical ensemble.)
\begin{expectation} (Almost continuous spectrum)
\label{continuousspectrum}
The gap between the energy eigenvalues of typical neighbouring high-energy eigenstates is $\Or[e^{-S}]$ in an interacting theory in the thermodynamic limit.
\end{expectation}
The motivation for this expectation is as follows.  Between the energy $[E-\Delta, E + \Delta]$ We expect to have $e^{S}$ states in a finite band of energies, $2 \Delta$. Except for an {\em exactly} free theory, interactions generically break all degeneracies. Therefore, the energy gap between neighbouring states scales like $e^{-S}$ in the thermodynamic limit. 

Expectation \eqref{continuousspectrum} also holds in theories with supersymmetry. Supersymmetry might ensure that some states, which saturate the BPS bound, are degenerate. However, as soon as we move slightly away from the BPS bound, the gap between eigenvalues becomes exponentially small. 

Some systems may have a forbidden-zone of energies in which states cannot exist. For example, superconformal field theories may have BPS representations that are separated from other representations with the same charges by a finite mass gap. (See, for example, the ``${\bf b}$'' representations in \cite{Kinney:2005ej}.) However, outside the forbidden zone, we again expect exponentially small gaps between neighbouring eigenvalues. 

Expectation \eqref{continuousspectrum} also holds in {\em integrable} systems. The statistical mechanics literature contains considerable discussion of the {\em statistics} of the distribution of energy eigenstates. The statistics of eigenvalues differ in integrable and chaotic systems (See, for example, \cite{casati1985energy,berry1977level}.) But the fact that the energy gap is $\Or[e^{-S}]$ holds almost universally.

In the context of the fuzzball proposal, there has been some discussion that the correct gap between energy-eigenstates, even at the supergravity point of the D1-D5 system, should be an inverse power of $\none \nfive$ (the product of the number of D1 and D5 branes) rather than an inverse exponential of this product \cite{Bena:2006kb,Tyukov:2017uig}. This is based on the fact that, at the orbifold point, the D1-D5 CFT does have a  gap that scales with ${1 \over \none \nfive}$. However, the orbifold CFT is a {\em free} theory. The moment we turn on the moduli that are necessary to reach the supergravity point, we expect that the degeneracies in the orbifold CFT will be destroyed. The entropy at energy $E$ scales as $S \propto \sqrt{\none \nfive E}$, and we expect that the gap between neighbouring energy eigenstates is of order $e^{-S}$ at a generic point in moduli space.

The exponentially small gap can be easily detected by a two-point function. For example, let $\al(t)$ be a simple operator localized in time. Then, given any typical high energy basis state, $|\fm \rangle$ of energy $E$ (which may {\em not} be an eigenstate), consider
\be
\label{smearedgreen}
G_{\smfunc}(\omega_0) =  \int d t \langle \fm | \al(t) \al(0) | \fm \rangle \smfunc_{\omega_0}(t) d t,
\ee
where $\smfunc_{\omega_0}(t)$ is a function whose Fourier transform is centered around $\omega_0$ with a width $\delta \gg E e^{-S}$. But  we can take $\delta$ to be very narrow. For example, in the D1-D5 theory, we may take $\delta = {1 \over (\none \nfive)^4}$ since this is still larger than $e^{-S}$. 

Then, by inserting a complete set of energy eigenstates, $|E_i \rangle$, 
\be
\label{completebasisgf}
G_{\smfunc}(\omega_0) = \sum_{i,j}  \langle \fm | E_i \rangle  \langle E_i | \al(0) | E_j \rangle \langle E_j| \al(0) | \fm \rangle \Big[\int \smfunc_{\omega_0}(t) e^{i (E_i - E_j) t} d t  \Big].
\ee
Since the difference $E_i - E_j$ takes on almost a continuous range of values we see that  $G_{\smfunc}(\omega_0)$ has support for a {\em continuous range} of  $\omega_0$.  Even if the state $|\fm \rangle$ is a supersymmetric state, we can choose an appropriate operator $\al$ that moves us off the BPS bound and whose two-point function displays a continuous spectrum.

States, where the two-point function does  not have a continuous spectrum for {\em any} simple operator typically correspond to microstates of a phase of zero-entropy. For such states, the three point function $\langle E_j | \al(0) | \fm \rangle$ that appears above vanishes for almost all except an exponentially small set of eigenstates, $|E_j \rangle$. For example, the boundary two-point function of light primary operators in the state dual to thermal AdS is expected to have a discrete spectrum. 

\begin{figure}[!h]
\begin{center}
\includegraphics[height=0.3\textheight]{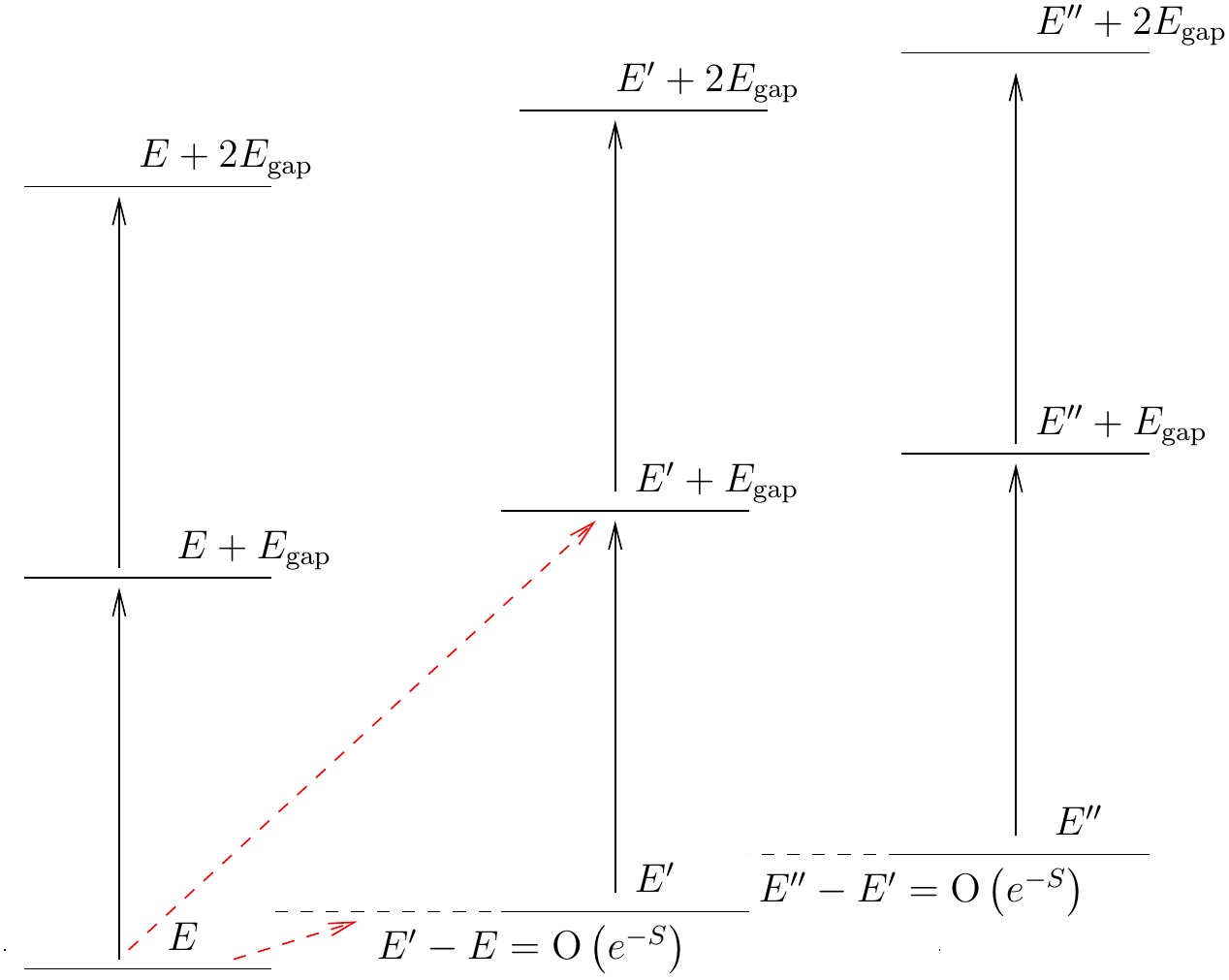}
\caption{\em An unusual possibility for the dynamics of fuzzball microstates. Probes of one microstate only excite other microstates in a single tower (solid black lines) and transitions between towers (dashed red lines) are disallowed. \label{towerfigure}}
\end{center}
\end{figure}

In the paper \cite{Bena:2018bbd}, it was argued that fuzzballs might represent typical states and still not show the continuous spectrum described above. The paper \cite{Bena:2018bbd} suggested that the matrix elements $\langle E_j | \al(0) | \fm \rangle$ could be subject to a {\em selection rule}: the matrix element vanishes unless $E_j - E = n E_{\text{gap}}$ where $E_{\text{gap}} \gg e^{-S}$ is some large gap and $n$ is an {\em integer}. Thus probing a particular fuzzball microstate with simple operators only excites a tower of integrally spaced excitations on top of that microstate. A probe of another microstate excites a parallel tower and it is impossible to move between towers by probing the system with simple operators. (See Figure \ref{towerfigure}.) The number of towers must be exponentially large to account for the total number of states.

This picture would suggest that the matrix elements between different states not only violate the eigenstate thermalization hypothesis (see below) but most matrix elements actually vanish.   Moreover, since it is impossible to transition between towers using simple operators, the system effectively breaks up into an exponentially large number of disconnected phases. If the D1-D5 system, or any gravitational system shows such {\em unusual} statistical behaviour, there should be a dynamical explanation for this. The selection rule above cannot just be {\em postulated} to save the fuzzball program from potential contradictions. In the absence of such a dynamical explanation, the simplest possibility is just that fuzzballs and other states where the two-point function does not show a continuous spectrum represent isolated states whose degeneracy is exponentially small compared to the set of all microstates.

\paragraph{\bf Relation to eigenstate thermalization.}
Before, we conclude our discussion on general statistical expectations, we should clarify the relation to the commonly discussed  {\em eigenstate thermalization hypothesis} (ETH)  \cite{srednicki1994chaos,srednicki1999approach}. 

The ETH  is usually stated for energy eigenstates. However, it really only relies on the assumption that, in a large statistical system, the eigenstates of most observables are likely to be highly scrambled versions of eigenstates of the Hamiltonian. So, given some basis of states for the microcanonical ensemble, $|\fm_i \rangle$, the ETH can be stated as
\be
\label{ethv1}
\langle \fm_j | \al | \fm_i \rangle = \al\big(E_i\big) \delta_{i j} + B\big({E_i + E_j \over 2}\big) e^{-S \over 2} R_{i j},
\ee
where R is a matrix of random phases and $E_i, E_j$ are the expectation values of the Hamiltonian in $|\fm_i \rangle, |\fm_j \rangle$ and $\al, B$ are smooth functions of their arguments. Note that $\al(E)$ is the microcanonical average of the observable at energy $E$. 

If the ETH holds for some basis of states, this implies that most elements of such a  basis are typical even for operators where the standard deviation is not parametrically small in the thermodynamic limit. 

The ETH is a significantly stronger assumption than the vanishing of the microcanonical standard deviation for classical observables that is an input to result \ref{thatypical}. The ETH  implies the typicality of most elements of the basis even for observables that have large $\sigmaens$. The ETH  arises from an assumption of quantum chaos, and in such a system expectation \ref{continuousspectrum} also holds. 

Therefore, the ETH is {\em stronger} than the assumptions of result \ref{thatypical} and expectation \ref{continuousspectrum}. Nevertheless, we expect the ETH to hold in theories with holographic duals \cite{Lashkari:2016vgj}. 

\subsection{Implications for the fuzzball program \label{subsecimplicationfuzz}}

Now we discuss several implications of these results  for the fuzzball program. In what follows, to make contact with the results above, we take the Hilbert space  $\hilbe$, to be the subspace that corresponds to black holes. If we consider large black holes in the AdS/CFT correspondence, then this subspace can simply be taken to be the microcanonical ensemble. However, note that the subspace corresponding to black holes exists even in flat space, where black holes do {\em not} dominate the canonical or microcanonical ensemble.

\paragraph{ Information-free nature of the horizon/stretched horizon.} The fuzzball program is often motivated by the idea  that the horizon should be replaced by the surface of a fuzzball that would contain ``information'' about the initial state. It is claimed that this structure would correct Hawking radiation at $\Or[1]$. 

We now apply result \ref{microtyp}. Let $\al$ correspond to an operator that measures  correlations  between different Hawking quanta.  For example, $\al$ may be a product of curvature invariants at distinct points. Since we are considering smooth geometries, such invariants are bounded and their fluctuations cannot be exponentially large. Then result \ref{microtyp} tells us that such observations in a typical microstate yield only an exponentially small amount of information. 

In particularly, it {\em cannot} be the case that Hawking radiation differs by $\Or[1]$ amounts between different typical microstates. 
It is sometimes claimed that ``high energy'' observables would take on a universal form but ``low energy'' observables at the scale of the Hawking radiation would differ between microstates \cite{Mathur:2012jk}. However result \ref{microtyp} allows no such freedom. {\em In a typical microstate, both high-energy and low-energy observables take on a universal value, and all features of the microstate can only be determined by exponentially precise observations.}

The idea that the surface of a black hole should contain ``information'' is often presented by making an analogy with a piece of coal. (For instance, see page 3 of \cite{Mathur:2008nj}.)  When coal burns, the properties of the outgoing radiation are strongly affected by the nature of its surface. However, this is a misleading analogy: everyday pieces of coal are {\em not} completely thermalized. They have a number of distinctive features because they are in highly {\em atypical} states.  A better example, to visualize a thermalized system, is a gas of radiation in a box. This gas is entirely featureless. Individual photons that emerge from the box contain almost no information about the state of the radiation inside the box; it is only by making exponentially precise measurements on the radiation that we can discern the state of the radiation.

\paragraph{ The universal fuzzball geometry.} 
Above, we argued that correlation functions of Hawking radiation measured in a typical microstate must take on a universal value.  We now argue that these correlators should correspond to correlators computed in effective field theory about an {\em approximately classical average bulk geometry} in the limit where the Planck length is smaller than all other scales. 

The signature of an approximately classical bulk geometry is that correlators of local operators factorize into products of lower-point functions \cite{ElShowk:2011ag}. 

In AdS/CFT the factorization of {\em boundary correlators} in the microcanonical/canonical ensemble can be proved at large-N using the standard factorization arguments. We can then use the standard HKLL construction \cite{Hamilton:2005ju,Hamilton:2006az,Hamilton:2006fh,Hamilton:2007wj} to construct approximately local operators and the factorization of boundary correlators implies the factorization of bulk correlators. This implies that, in AdS/CFT,  the microcanonical/canonical ensemble is dual to an approximately classical bulk geometry. By result \ref{microtyp} this is also the geometry dual to a typical microstate. This geometry can also be used to compute  n-point functions of simple operators to excellent accuracy.

In flat space, we cannot make such a clear argument that averages computed in $\hilbe$ correspond to an approximately classical bulk. But, even here we expect that S-matrix elements will factorize if the Planck length is much smaller than other length scales in the problem. These S-matrix elements can be used to reconstruct a bulk geometry that is approximately classical.

A priori, we do not know what this universal classical geometry should be. In section \ref{sectwocharge}, we will {\em compute} this average geometry for the two-charge Lunin-Mathur solutions that have been quantized. However, in more general settings that correspond to large black holes, we cannot compute this average geometry since all fuzzball solutions have neither been found nor quantized. So, in the remainder of this paper, we will simply proceed with the following expectation.
\begin{expectation}({\em Conventional geometry as average})
\label{euclideanaverage}
The conventional black-hole geometry --- after incorporating classical string-theory corrections ---  correctly computes the average value of bulk observables such as the metric and correlation functions of the metric as long as we are more than Planck length outside the horizon.
\end{expectation}

We believe that this is a fairly uncontroversial assumption.  If the geometry obtained by averaging over all microstates differs significantly from the black-hole geometry, this has significant implications for AdS/CFT: it would imply the computations in a thermal state in the CFT should be matched to bulk computations in this special average fuzzball geometry (whatever it may be) rather than the black hole. This would be the case even for time-ordered correlators that are obtained naturally from {\em Euclidean computations}. Therefore any claim that expectation \ref{euclideanaverage} is violated must be accompanied by an explanation for why the Euclidean saddle point is not adequate for correlators outside the horizon.  We are not aware of any place in the literature where such a strong claim has been made.

Expectation \ref{euclideanaverage} allows for the possibility that the average geometry to have Planck-scale deviations from the conventional geometry. We discuss these deviations in greater detail below.
 
\paragraph{Distinct fuzzballs as a basis?}
Result \eqref{microtyp} implies that the geometries corresponding to {\em typical} states can only differ by an exponentially small amount from the average geometry,
We certainly do not expect to represent such exponentially small deviations in terms of a classical metric, and therefore the idea that fuzzballs can represent typical microstates is entirely untenable. Typical microstates are represented by the {\em same} average geometry.

One might imagine that while it is impossible to describe the different typical microstates using geometries, perhaps one could use a set of distinguishable geometries as a {\em basis} for all microstates of the black hole. 
However, we will now show that result \ref{thatypical}, together with expectation \ref{euclideanaverage} constrains how much the typical element of the basis can differ from the conventional black hole. To make this precise, we pause to define some useful intermediate quantities that we will use later in the paper as well.

\paragraph{ The ``difference'' and ``quantumness'' parameters. }
Let $\opbulk(r)$ be a simple {\em bulk} observable.  For example,  $\opbulk(r)$ may be  some coordinate invariant function of the metric. Here $r$ denotes the ``radial'' coordinate in a coordinate system where the horizon is at $r = r_h$ and $r = \infty$ is the asymptotic region.  To make physical meaningful comparisons, $r$ should be defined through the physical area of a compact submanifold in the geometry.

Let $\opbulkbh(r)$ be the expectation value of this observable in the black-hole,  For a fuzzball microstate, $|\fm \rangle$, we denote
\be
\langle \fm | \opbulk(r) | \fm \rangle = \opbulkfuzz(r).
\ee 
The quantum fluctuations of this operator, in the fuzzball state, are measured by
\be
\sigma^2(r) = \langle \fm | \opbulk(r)^2 | \fm \rangle - \langle \fm | \opbulk(r) | \fm \rangle^2,
\ee
where the product of operators at a point may need to suitably renormalized. 

We now define two parameters.  The {\em difference parameter}, $\devpar$ is defined as
\be
\label{diffparamdef}
\devpar_{\op}(r) = \big|{\opbulkbh(r) - \opbulkfuzz(r)  \over \opbulkfuzz(r)}\big|.
\ee
The {\em quantumness parameter}, $\quantpar$ is defined as
\be
\label{quantparamdef}
\quantpar_{\op}(r) = \big| { \sigma(r) \over \opbulkfuzz(r)} \big|.
\ee

For a classical solution to be ``interesting'' we require that the difference parameter be large. On the other hand, for the classical solution to be reliable, the quantumness parameter must be parametrically suppressed. This is particular important in a non-linear theory like gravity. It makes no sense to trust classical general relativity in a regime where quantum fluctuations of the metric are of the same order as the metric itself. 

We argue below that typical fuzzballs cannot meet both conditions simultaneously. In the region where they are interesting, they also become unreliable.

\paragraph{Deviations of individual fuzzballs from the average geometry.}
From result \ref{microtyp}, the fluctuations that enter result \ref{thatypical} are the same as quantum fluctuations in a typical state. Since we argued above that typical states correspond to the conventional black-hole geometry, we can estimate the fluctuations that enter result \ref{thatypical} by estimating quantum fluctuations in the black-hole geometry.
\be
\begin{split}
\sigmaens(r) &= {1 \over e^{S}} \tr_{\hilbe} \opbulk(r)^2 - \left({1 \over e^{S}} \tr_{\hilbe} \opbulk(r) \right)^2 \\
&=  \langle \Psi | \opbulk(r)^2 | \Psi \rangle - \langle \Psi | \opbulk(r) | \Psi \rangle^2 + \Or[e^{-{S \over 2}}],
\end{split}
\ee
where $|\Psi \rangle$ is a typical microstate,  and we have used result \ref{microtyp} in the second equality. Note that $\sigmaens(r)$ may {\em not} coincide with $\sigma(r)$ defined above if $|\fm \rangle$ is not a typical state.

The leading quantum fluctuations in the black-hole geometry appear with a factor of ${1 \over G_N}$ and on dimensional grounds, we expect that they are proportional to $\big({\ell \over \lpl}\big)^{d-2}$ where $\ell$ is the typical length scale in the geometry. If we are far away from the horizon, then we expect that $\ell \leq r_h$. We also note that  the entropy is proportional to 
$\left({r_h \over \lpl} \right)^{d -2 }$.  Therefore for simple gauge-invariant observables made out of the metric, we expect that for observables with a non-zero classical expectation value\footnote{What we will need, in subsequent sections, is just that ${\sigmaens(r) \over \opbulkbh(r)}$ is small --- not that it takes the precise value predicted by the black-hole geometry. This may hold even  if expectation \ref{euclideanaverage} fails: as long as the typical microstate in $\hilbe$ corresponds to an approximately classical geometry, we can estimate $\sigmaens$ by quantizing metric fluctuations in this geometry, and these quantum fluctuations will be small compared provided that typical curvatures are small.}
\be
\label{smallsigmaens}
{\sigmaens^2(r) \over (\opbulkbh(r))^2} = \Or[{1 \over S}], \quad r - r_h \gg \lpl.
\ee
(The reader may consult \cite{Candelas:1980zt} for a concrete calculation of quantum fluctuations in the black-hole background.)

However, then result \ref{thatypical} tells us that for all but a vanishing fraction of fuzzball states, we also have
\be
\label{smalldevawayhor}
\devpar_{\op}(r) =  \big|{\opbulkbh(r) - \opbulkfuzz(r) \over \opbulkfuzz(r)}  \big| = \Or[{1 \over \sqrt{S}}], \quad r - r_h \gg \lpl.
\ee
Moreover, if $\devpar_{\op}$ is very small then quantum fluctuations of the metric in the fuzzball geometry are also very close to quantum fluctuations in the black-hole geometry. Therefore 
\be
\quantpar_{\op}(r) = \Or[{1 \over \sqrt{S}}].
\ee
So the deviation of the fuzzball metric from the black-hole metric can {\em at most} be of the same order as the quantum fluctuations of the metric. 

It is important that \eqref{smalldevawayhor} continues to hold when $r = r_h + \ls$, where $\ls$ is the string-length. The black hole metric is corrected at the string-scale but we can compute fluctuations of the metric, using Euclidean quantum gravity, and we do  {\em not} expect quantum fluctuations in the black-hole geometry to become significant at the string-scale. 

Since, by definition, fuzzballs have no horizon  they must start to deviate appreciably from the conventional black-hole geometry at some point. The argument above tells us that for typical fuzzballs, this can only happen when  $r - r_h = \Or[\lpl]$.  This is precisely where expectation \ref{euclideanaverage} also allows the average geometry to deviate from the conventional geometry.\footnote{Standard calculations in the conventional black-hole geometry suggest that when the geometry has a macroscopic horizon, we do {\em not} expect any unusual effects in the near-horizon region and $\sigmaens$ continues to be small there.  However, it is difficult to {\em prove} this even in holography, since the HKLL construction requires very long time-bands on the boundary to represent physics in the region $r - r_h = \Or[\lpl]$. The length of these bands scales with N and may interfere with standard large-N counting. 
So, in this paper, we make a generous assumption for the fuzzball program by allowing the possibility that some unknown hitherto unknown effect invalidates the standard calculation of $\sigmaens$ within a Planck length of  the horizon and somehow makes it large. }

But this means that the geometric solution ---  corresponding to a typical basis state or the average geometry ---  must explicitly have {\em Planck-scale} structures, presumably through an explicit length-scale that takes on a Planck-scale value. However, we expect that any length-scale in quantum gravity will itself undergo fluctuations of the size of the Planck-scale. Therefore, in the region where we are very close to the horizon, if the fuzzball has explicit Planck scale features, then quantum fluctuations in the metric are expected to be of the same order as these Planck-scale structures. So, 
\be
\label{farhorizon}
\devpar_{\op}(r) = \Or[1],~~\text{but}~~\quantpar_{\op}(r) = \Or[1], \quad \text{when}~r - r_h = \Or[\lpl].
\ee

But if the parameter $\quantpar_{\op}=\Or[1]$, then the classical solution becomes completely unreliable. So, if we explicitly insert Planck-scale features into the fuzzball solution in order to satisfy result \ref{thatypical}, then we run into the difficulty that the geometry becomes unreliable just where it appears to be interesting.

\begin{figure}
\begin{center}
\includegraphics[width=0.8\textwidth]{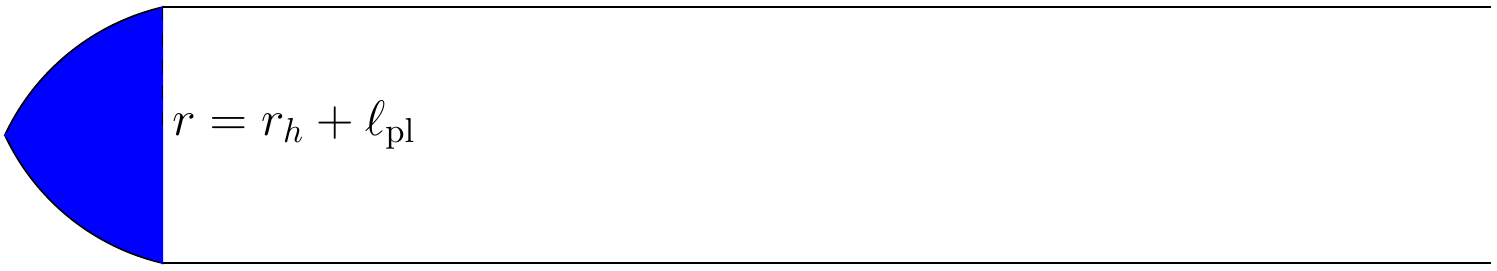}
\caption{\em A schematic representation of what a typical fuzzball geometry must look like, if fuzzballs represent black-hole microstates. The geometry must closely resemble the black-hole geometry away from the horizon (unshaded region) and then suddenly deviate away to cause some extra-dimension to pinch off when we reach within Planck length of the horizon (blue region).  \label{fuzzplanck}}
\end{center}
\end{figure}

To summarize, we have argued the following. If fuzzballs are to represent typical microstates then they must have the following features:
\begin{enumerate}
\item
When we are far away from the horizon (in Planck units), the fuzzball geometry is indistinguishable from the black-hole geometry up to terms that are suppressed by the black-hole entropy. This follows from the fact that the {\em average} fluctuations of the metric --- which can be computed in the black-hole geometry --- are small, and then result \ref{thatypical} limits the extent to which  typical basis elements can differ from the average.
\item
When we approach within a Planck length of the horizon, the fuzzball geometry may appear to deviate from the conventional black hole. But such a geometry must explicitly contain Planckian structures, and then we expect that quantum fluctuations will become large  and so the fuzzball geometry becomes unreliable. (Note that this is contrast to the conventional black-hole geometry, which remains perfectly reliable close to the horizon.)
\end{enumerate}
This picture of the fuzzball is shown schematically in Figure \ref{fuzzplanck}

Therefore, {\em it is wrong to think of fuzzballs as macroscopically distinct geometries, which somehow average out to give the same answer as the black hole.}  Rather, typical fuzzballs  must all look like Figure \ref{fuzzplanck} to satisfy result \ref{thatypical}. Fuzzballs which have structure on a scale larger than the Planck scale  can only be a vanishing fraction of microstates by result \ref{thatypical}.

\paragraph{ Requirement of large red-shifts.} 
The discussion above utilizes expectation \ref{euclideanaverage}, which is what leads to equation \eqref{smallsigmaens}. However, even without invoking expectation \ref{euclideanaverage},  we can still use  expectation \ref{continuousspectrum} to justify the important aspects of the picture shown in Figure \ref{fuzzplanck}. 

Consider a  quantum field that propagates in the bulk whose excitations about the vacuum are gapped. In global AdS we can consider a massless field. If we are considering flat space or Poincare AdS, we can consider a massive field. Then expectation \ref{continuousspectrum} implies that the {\em asymptotic} two-point function of this field must be supported at arbitrary frequencies even though, locally, field excitations are gapped.

The reason that  black holes allow this phenomenon is because of the infinite redshift at the horizon. This red-shift allows for arbitrary low-energy excitations. This is {\em not} a bug; it is a {\em feature} of the black-hole geometry which ensures that it can be interpreted as a heavy state in a quantum-mechanical system with large entropy.

If fuzzballs are to represent black-hole microstates, they must also support a continuous spectrum. Therefore, the fuzzball geometry must also have an extremely large-shift. In particular, {\em if the geometry caps off to form a fuzzball at any length-scale that is visible classically, then the inverse of this length-scale will be visible as an energy gap that would violate expectation \ref{continuousspectrum}.}  Once again we see that the requirement of an almost continuous spectrum disallows fuzzballs that are of size $\ls$ or any other classical length-scale. 

What if the geometry caps  off at $r = r_h + \lpl$? Even such a geometry would not support the exponentially suppressed gap that is required around a heavy pure state since it would allow, at most, an energy gap that is power-law suppressed in the entropy. So the only possibility is for the geometry to stop making sense classically below $r =r_h + \lpl$. However if this happens we return to our conclusion above, displayed in Figure \ref{fuzzplanck}:  fuzzballs are uninteresting in most of space (where $r - r_h \gg \lpl$) and unreliable where they are interesting ($r - r_h = \Or[\lpl]$). 

Larger fuzzballs that are both reliable and interesting by virtue of having larger than Planck scale structure are {\em irrelevant} to the discussion of black-hole microstates since they do not have the right energy gap expected in a system with large entropy.

\paragraph{Eigenstate thermalization.}
The arguments that led to the structure expected from typical fuzzball geometries shown in Figure \ref{fuzzplanck} assumed that the bulk metric was a good 
observable with small quantum fluctuations. We believe that this is a very robust assumption. 

However, if we assume the ETH, then we can deduce such a structure for typical fuzzballs while restricting our discussion to only asymptotic
observables. If we apply the ETH in the form \eqref{ethv1} to fuzzball states, then we expect to get the microcanonical average
\be
\langle \fm | \al |\fm \rangle = \al(E),
\ee
where $E$ is the energy of the fuzzball state and $\al(E)$ is the microcanonical average of $\al$. 

Therefore, the ETH tells us that even for asymptotic observables, we should get precisely the same value in a typical fuzzball microstate as we do in the conventional black hole. 

Intuitively, this rules out fuzzballs that differ at leading order from the black-hole metric. This is because if the geometry differs at leading order, then a simple scattering experiment with waves sent in from asymptotic infinity will detect this variation and produce an answer that fails to satisfy the ETH. We will see in section \ref{secmulticharge} that this is precisely what happens for fuzzballs with macroscopic structures.

\paragraph{A cautionary note.} We close this subsection with a note of caution. The reader will note that our arguments above have been based on simple physical expectations and general results from statistical mechanics. This makes them broadly applicable but it also means that these arguments are only suggestive of the difficulties that the fuzzball program must surmount and cannot be taken as a proof that the program is not viable.  

In section \ref{sectwocharge} and \ref{secmulticharge} we will verify the correctness of these arguments in specific examples. But if the fuzzball program is to be carried through to completion in any system, this must involve a loophole in the arguments outlined above, and it would be interesting to understand the origins of such a loophole. 
 
\subsection{Indirect arguments for horizon structure  \label{subsecindirectarg}}

In an attempt to sidestep arguments of the kind that we have provided above, Mathur put forward indirect arguments to show that the horizon must have structure \cite{Mathur:2009hf}. Mathur's argument was based on the strong subadditivity and it was later used to produce the firewall paradox \cite{Almheiri:2012rt}.

This argument has been reviewed several times and, in particular, we refer the reader to section 6.1 of \cite{Ghosh:2017pel} for an up-to-date review. Very briefly, the idea is to divide a nice slice in the late-part of an evaporating black hole into a near horizon region outside the horizon, $B$, the rest of the region outside the black hole, $A$, and a near-horizon region inside the horizon, $C$. 

An argument due to Page \cite{Page:1993df} suggests that in the latter part of black-hole evaporation, $S_{AB} < S_{A}$, whereas the smoothness of the interior suggests that $S_{BC} < S_{C}$. These two inequalities are in contradiction with the strong-subadditivity of entropy $S_{AB} + S_{BC} \geq S_{A} + S_{C}$. 

In \cite{Mathur:2009hf}, this was used to argue that we should drop $S_{BC} < S_{C}$, and consequently that we should drop the idea that the interior of the black hole is a smooth featureless region.

The difficulty with this argument is that the strong-subadditivity of entropy {\em assumes} that the set of observables on a nice slice factorize into observables on $A, B,C$. However, in quantum-gravity this is simply not true. To the contrary, observables in $C$ can be written as scrambled versions of observables in $A$.

We emphasize that this is {\em not} some hand-waving argument about complementarity. This phenomenon --- where local operators in one region can be rewritten as complicated local operators in another region --- can be made absolutely precise in empty AdS \cite{Banerjee:2016mhh}. Moreover, it was shown in \cite{Ghosh:2017pel} that if we take strong-subadditivity seriously in quantum gravity, we can construct paradoxes even in empty AdS (See section 6.4 of \cite{Ghosh:2017pel}.)

Even in flat space, where the explicit formulas of \cite{Banerjee:2016mhh} cannot be written down, we have substantial evidence from the dynamical breakdown of string perturbation theory in some regimes that a similar loss of locality occurs at a non-perturbative level in quantum gravity \cite{Ghosh:2016fvm,Ghosh:2017pel}. 

The arguments of \cite{Mathur:2009hf} were preceded by other arguments that the information paradox requires a smooth horizon to be replaced by a geometry with details that would carry information about the initial state. These arguments are unsatisfactory because it is well understood that information can be stored in exponentially small correlations between different Hawking quanta \cite{Papadodimas:2012aq}. So, the recovery of information outside the black hole does not, by itself, require any large modifications to the classical black-hole geometry.

 A related set of ideas suggests that a pure-state cannot have a horizon, because a horizon has entropy whereas pure states cannot have entropy \cite{Skenderis:2008qn}. However, this is incorrect. The thermodynamic entropy should not be conflated with the von Neumann entropy. So, pure-states can also have thermodynamic entropy which arises after we coarse-grain the system and this entropy reflects the fact that coarse-grained probes of the system leave its fine-grained features undetermined \cite{jaynes1957information,jaynes1957informationII}. In a theory of quantum gravity, the geometry is a tool to encode expectation values of the metric and its low-point correlators --- it is thus an explicitly coarse-grained probe of the full theory. So it is perfectly consistent for the geometry to be described by a metric with a horizon whose entropy reflects the fact that we have ignored the fine-grained {\em non-metric} degrees of freedom that are part of the full description. 

We note, parenthetically, that in anti-de Sitter space, the papers \cite{Almheiri:2013hfa,Marolf:2013dba} made an entirely independent set of arguments to suggest that the black-hole interior cannot be represented in the boundary CFT. These arguments  are relevant for large AdS black holes that are thermodynamically stable. Moreover, even if they are correct, they suggest that the black hole has a horizon and the interior of a black hole has a firewall rather than a fuzzball.   For this reason the arguments of \cite{Almheiri:2013hfa,Marolf:2013dba}  are not directly relevant here,  and a detailed discussion of their merits is beyond the scope of this paper.  Nevertheless, a brief summary of their status is as follows.

Several authors \cite{Papadodimas:2013wnh,Papadodimas:2013jku,Papadodimas:2015xma,Papadodimas:2015jra,Verlinde:2013uja,Verlinde:2013qya,Verlinde:2012cy, Guica:2014dfa} have pointed out that interior operators can be constructed using a suitably state-dependent construction.  The authors of \cite{Harlow:2014yoa,Marolf:2015dia} suggested that state-dependence would lead to observable peculiarities for an infalling observer but it was explained in \cite{Raju:2016vsu} that these effects were not observable in physically reasonable experiments. 

Physically, our understanding of the origins of state-dependence has also advanced. The state-dependence of the interior can be understood as arising from a fat-tail in the inner-product of coherent states in gravity \cite{Papadodimas:2015xma,Papadodimas:2015jra} and this fat-tail also contributes to the fact that interior operators, when gauge-fixed in a particular manner, may fail to satisfy a non-perturbative version of the Hamiltonian constraints \cite{Jafferis:2017tiu}. The origins of state-dependence can also be studied in toy-models \cite{Bzowski:2018aiq}. Of course, several questions about state-dependence and the reconstruction of the black-hole interior in AdS/CFT remain to be understood.

To summarize this subsection, we have argued that indirect arguments for the relevance of fuzzballs to black-holes are invalid. This is an important point. It shows that one cannot concede the limitations of supergravity solutions --- as is sometimes done in the fuzzball program --- but yet argue that black-holes do not have smooth interiors. {\em Neither a resolution to the information paradox nor an understanding of the black-hole entropy requires the existence of fuzzballs.}  The relevance, or lack thereof, of fuzzballs to the study of black-holes must follow from a study of the known fuzzball solutions. If these solutions are irrelevant to black-holes, there is {\em no other valid argument for the relevance of fuzzballs} for black-holes.

\section{Quantum aspects of  the two-charge solutions  \label{sectwocharge}}
In this section, we examine the original two-charge fuzzball solutions that were discovered in \cite{Lunin:2001fv,Lunin:2002bj}. The literature on these solutions, and their relation to CFT microstates is extensive. Our analysis will be simple and independent, but we note a few salient results in the literature. In \cite{Sen:2009bm} and then in \cite{Chen:2014loa}, it was pointed out that the supergravity solution was not valid for typical states and as one travels towards the fuzzball cap, it is necessary to transition out of the D1-D5 duality frame. This can indeed be done in some cases, and for specific solutions the full stringy description was analyzed in  \cite{Martinec:2018nco}. We will reach a similar conclusion, although our reasoning will be slightly different and placed within the framework developed in section \ref{secstatprelim}.  

There has also been work on identifying specific solutions with  microstates in the orbifold CFT \cite{Skenderis:2006ah,Kanitscheider:2006zf,Kanitscheider:2007wq}. We note that even for very simple states, such an identification must be performed carefully since the supergravity point is very far from the orbifold point in the D1-D5 moduli space and, moreover, states  {\em cannot} be uniquely identified just by specifying one-point functions of a few operators. In fact, in general, such an identification is impossible since the matching between states at different points in moduli-space is path-dependent \cite{deBoer:2008qe}. Holographic correlators have also been calculated in these solutions \cite{Bombini:2017sge,Galliani:2017jlg,Galliani:2016cai} as we will do for multi-charge solutions in section \ref{secmulticharge}.

For us, what is important, is that these solutions were quantized in \cite{Rychkov:2005ji}, following a suggestion made in \cite{Lunin:2002iz}. Therefore, we can study the quantum mechanics of this set of solutions and we will use this system to {\em verify} the arguments of section \ref{secstatprelim}. We will compute the {\em average} fuzzball geometry, and we will also compute {\em quantum fluctuations} in this geometry. This allows us to compute the parameters $\devpar$ (defined in \eqref{diffparamdef}) and $\quantpar$ (defined in \eqref{quantparamdef}). This system differs slightly from the setup of section \ref{secstatprelim} because the horizon of the conventional solution is of zero size because a circle in the geometry shrinks to zero at that point. Nevertheless --- using the  size of this circle as a  measure of the distance from this zero-size horizon -- we find that
\begin{enumerate}
\item
As the distance from the position of the conventional horizon becomes greater than the Planck length, the average fuzzball geometry tends rapidly to the classical geometry.
\item
The average geometry starts deviating from the conventional geometry when we are within Planck length, and {\em not} string length, of the horizon. Moreover,  most of the entropy of the set of solutions comes from solutions that differ from the conventional geometry at the Planck scale.
\item
In the region where deviations of the average geometry from the conventional geometry are appreciable, quantum fluctuations are of the same order as the expectation values of components of the metric. Therefore the solution is entirely unreliable.
\end{enumerate}

We will consider the two-charge solutions  in the following form, using the conventions of  \cite{Rychkov:2005ji}.
\be
\label{fuzzmetric}
\begin{split}
&ds^2 = e^{-{\phi \over 2}} ds_{\text{str}}^2;\quad e^{-2 \phi} = {f_5 \over f_1}; \\ 
&ds_{\text{str}}^2 = {1 \over \sqrt{f_1 f_5}} \left(-(d t + A)^2 + (d y + B)^2 \right) + \sqrt{f_1 f_5} d \vec{x}^2 + \sqrt{f_1 \over f_5} d \vec{z}^2; \\ 
&f_5 = 1 + {Q_5 \over L} \int_0^L {d s \over | \vec{x} - \vec{F}(s)|^2}; \quad 
f_1 = 1 + {Q_5 \over L} \int_0^L {|\vec{F}'(s)|^2 \over |\vec{x} - \vec{F}(s)|^2}; \\
&A_i= {Q_5 \over L} d x^i \int_{0}^L {F_i' (s) \over |\vec{x} - \vec{F}(s)|^2} d s; \quad
d B = *_4 d A; \\
&C = {1 \over f_1} (d t + A) \wedge (d y + B) + {\cal C}; \quad d {\cal C} = -*_4 d f_5.
\end{split}
\ee
Here $\vec{z}$ denotes four compact directions. The conventional solution is obtained simply by setting
\be
\label{convensubst}
f_1 \rightarrow 1 + {Q_1 \over \vec{x}^2}; \quad f_5 \rightarrow 1 + {Q_5 \over \vec{x}^2}.
\ee
and setting $A = 0, B = 0$. 

These solutions can be systematically quantized by recognizing that the space of classical solutions can be bijectively mapped to points on the phase space; the action of the theory yields a symplectic form on this space, and the machinery of geometric quantization can then be applied to obtain a Hilbert space \cite{dedecker1953cvf,crnkovic1987cdc,zuckerman1987apa}. The result of this process is very simple. The quantization promotes the functions $F^k(s)$ to operators as follows
\be
F^k(s) = \mu \sum_{n > 0} {1 \over \sqrt{2 n}} \left(a^k_n e^{-2 \pi i n s \over L } + (a^k_n)^{\dagger} e^{2 \pi i n s \over L}\right),
\ee
where $[a_n, a_m^{\dagger}] = \delta_{n m}$.  The various parameters that appear here are defined as
\be
\mu = {g_s \over R \sqrt{V_4}}; \quad L = {2 \pi Q_5 \over R}.
\ee
Here $R$ is the coordinate radius of the $y$-direction and $V_4$ is the coordinate volume of the compact manifold. These are moduli of the solution. We are working in units where the {\em string length is set to unity.} The charges are related to the brane-numbers by
\be
\label{chargebrane}
Q_5 = g_s \nfive; \quad Q_1 = {g_s \none \over V_4}.
\ee

For the purposes of counting states, it will be useful to define the following ``Hamiltonian''
\be
H = \sum_{n > 0, k} n  (a^k_n)^{\dagger} a^k_n,
\ee
 where we have a infinite set of harmonic oscillators with creation and annihilation operators specified by $a^k_n$ and $k$ runs over $1 \ldots 4$.  The fuzzball states dual to the D1-D5 system with charges ($Q_1$, $Q_5$) are defined to be the states in this quantum system that have $H = \none \nfive$. 

We will not attempt to compute the full quantum expectation value of the metric. Instead, we will focus on the following list of quantum expectation values,
\be
\typexp[f_5 - 1], \quad \typexp[f_1 - 1], \quad \typexp[A_i],
\ee
in a typical state, $|\Psi \rangle$. Here,  ``typical state'' is used in the sense of result \ref{microtyp}. These one-point expectation values were also calculated in \cite{Balasubramanian:2008da}, and our results agree precisely with theirs.  We will not consider $B_i$ separately since this field is defined through the dual of $A$. Note that we also subtract off the uninteresting $1$ in both $f_1$ and $f_5$.  

$A_i$ vanishes in the conventional geometry, and it will turn out that $\typexp[A_i]$ also vanishes. So the difference and quantumness parameters (\eqref{diffparamdef} and \eqref{quantparamdef}) are not well defined for this observable. Therefore, we will consider another one-form that does not appear in the metric but is also an interesting probe of the geometry
\be
\wg_i= {Q_5 \over L} \int_{0}^L {F_i(s) \over |\vec{x} - \vec{F}(s)|^2} d s.
\ee
This quantity is of interest since it vanishes in the conventional geometry but, as we will find, $\typexp[W_i]$  takes on a finite value. So one can ask if this finite value is reliable.

We will also compute the quantum fluctuations in these quantities by computing  the following quantum two-point functions
\be
\typexp[(f_5 - 1)^2], \quad   \typexp[(f_1 - 1)^2] ,  \quad  \typexp[ \wg_i \wg_j].
\ee
We will use these two-point functions to evaluate the difference and quantumness parameters for these observables. These calculations will allow us to  verify all the expectations outlined in section \ref{secstatprelim} in a precise setting.

\subsection{One-point functions: $\typexp[f_5 - 1], \typexp[f_1 - 1], \typexp[A_i], \typexp[\wg_i]$ \label{subseconept}}

Using result \ref{microtyp}, the expectation value of the observables above in a typical state can be computed by considering the microcanonical trace. Therefore, we can consider the generating function
\be
\expgenfn[\chi, \alpha] = {1 \over e^{S(E)}} \tr_{\text{mic}} \left[ \int_{-\infty}^{\infty} \prod_k d g_k \int_0^{\infty} d t \int_0^L {d s \over L} :e^{\exponfn}:  \left({t \over \pi}\right)^{2}  \right],
\ee
where
\be
\exponfn = -\sum_k t g_k g_k  + 2 i t (x^k - F^k(s)) g_k +  \sum_j \alpha^j F^j(s) + \chi^j {d F^j(s) \over d s}.
\ee
Here the trace is taken over all energy eigenstates of the Hamiltonian with a large total energy, 
\be
E  = \none \nfive,
\ee
and the degeneracy of states at that energy is given by $e^{S(E)}$. By normal ordering we mean that when  we expand the exponential in terms of creation and annihilation operators, we move all annihilation operators to the right. This, of course, involves a necessary choice of how to interpret the quantum operator corresponding to the classical quantity.

From result \ref{microtyp}, in a typical microstate $|\Psi \rangle$, we expect
\be
\label{typexps}
\begin{split}
&\typexp[f_5] = 1 + Q_5 \expgenfn[\chi = 0, \alpha = 0]; \\
&\typexp[f_1]= 1 +  Q_5 \lim_{\chi \rightarrow 0} {\partial \over \partial \chi^i} {\partial \over \partial \chi^i} \expgenfn[\chi, \alpha = 0]; \\
&\typexp[A_i] = Q_5 \lim_{\chi \rightarrow 0} {\partial \over \partial \chi^i} \expgenfn[\chi, \alpha=0]; \\
&\typexp[\wg_i] = Q_5 \lim_{\alpha \rightarrow 0} {\partial \over \partial \alpha^i} \expgenfn[\chi = 0, \alpha]. \\
\end{split}
\ee 

We can equivalently compute this object by using the equivalence of the canonical and microcanonical ensemble at large $S(E)$
\be
\expgenfnbeta[\chi, \alpha] = {1 \over Z(\beta)}  \tr \left[ e^{-\beta H} \int_{-\infty}^{\infty} \prod_k d g_k \int_0^{\infty} d t \int_0^L {d s \over L} :e^{\exponfn}:  \left({t \over \pi}\right)^{2}  \right] ,
\ee
where the ``temperature'', $\beta^{-1}$, is set by demanding that the expectation value of the Hamiltonian be $\none \nfive$ and $Z(\beta) = \tr(e^{-\beta H})$. The equivalence of ensembles implies that 
\be
\expgenfnbeta[\chi, \alpha] = \expgenfn[\chi, \alpha] + \Or[{1 \over \sqrt{S(E)}}],
\ee
and this accuracy is sufficient for our purpose. Similarly, the thermal expectation values, $\thermexp[f_5], \thermexp[f_1], \thermexp[\wg_i]$ obtained from this generating function match the typical expectation values of \eqref{typexps} up to terms suppressed by the entropy.

Before evaluating the traces we need, we remind the reader of a few simple results. If we consider a {\em single} simple harmonic oscillator, corresponding to a given value of $n$ and $k$, then in a number eigenstate of that oscillator,  $|N^k_n \rangle$, for any values of the c-number coefficients $c_n^k$ and $d_n^k$, we have
\be
\langle N^k_n | e^{c_n^k (a_n^k)^{\dagger}} e^{d_n^k a_n^k} | N^k_n \rangle = \sum_{r=0}^N {N! \over (N-r)! (r!)^2}  (c_n^k d_n^k)^r = \sum_{r = 0}^{\infty} {N^k_n (N^k_n-1) \ldots (N^k_n - r + 1) \over (r!)^2} (c_n^k d_n^k)^r.
\ee
Here in the last step, we have simply noted that the sum over $r$ can be extended till $\infty$. All terms larger than $N$ vanish because of the factor of $(N-r+1)$ in the numerator.  If we take the thermal trace and denote $z = e^{-\beta}$, we find
\be
\begin{split}
\tr\left(e^{-\beta n N^k_n} e^{c_n^k (a_n^k)^{\dagger}} e^{d_n^k a_n^k} \right) &=  \sum_{N_{n}^k=0}^{\infty}  \sum_{r=0}^{\infty} {N_n^k (N_n^k - 1) \ldots (N_n^k-r+1) \over (r!)^2} z^{n N_n^k} \left(c_n^k d_n^k \right)^{r} \\
&= {1 \over 1 - z^n}\exp\big({c_n^k d_n^k z^n  \over 1 - z^n} \big),
\end{split}
\ee
where we have used the identity 
\be
\sum_{N=0}^{\infty} N (N-1) \ldots (N-r+1) {x^N} = r! {x^r  \over (1 - x)^{r+1}}.
\ee

We now note that
\be
\begin{split}
\exponfn =  & \sum_{k,n} \frac{\mu  e^{-\frac{2 i \pi  n s}{L}} a_n^k (-2 i  t g_k+  \alpha _k- i {2 \pi  n \over L} \chi _k)}{\sqrt{2 n} L }+\frac{\mu  e^{\frac{2 i \pi  n s}{L}} (a_n^k)^{\dagger} (-2 i  t g_k+  \alpha_k + i {2 \pi  n \over L} \chi_k)}{\sqrt{2 n}  } \\ &- \sum_k t g_k (g_k-2 i x_k).
\end{split}
\ee
The expression above is just in the form we need. We see that the thermal trace breaks up into a product of the traces over individual oscillator sectors, and moreover that for each oscillator the coefficients $c_n^k$ and $d_n^k$ can be identified from the expression above.
This leads to
\be
\label{thermaltraceexpr}
\begin{split}
\tr(e^{-\beta H} :e^{\exponfn}:) &= \exp\Bigg[\sum_k 2 i t g_k x_k-t g_k^2 + \sum_n \log(1 - z^n) \\ &+ \sum_{n}{1 \over 1 - z^n}\big( \frac{-2 \mu ^2 t^2 g_k^2 z^n}{n }-\frac{2 i \mu ^2 t g_k \alpha _k z^n}{n }+\frac{2 \pi ^2
   \mu ^2 n \chi _k^2 z^n}{L^2 }
+\frac{\mu ^2 \alpha _k^2 z^n}{2 n }\big)\Bigg].
\end{split}
\ee

Note that if we take the limit $t, g_k, \alpha_k, \chi_k \rightarrow 0$ in 
\eqref{thermaltraceexpr}, we simply get the partition function, which is
\be
Z(\beta) =  e^{\sum_{n,k} \log(1 - {z^n})}.
\ee
We can expand the logarithm in a power series and the interchange the order of sums to get
\be
Z(\beta) = \exp\Big[d {\sum_{n,m} {1 \over m} z^{n m}}\Big] = \exp\Big[d {\sum_{m} {z^{m } \over m(1 - z^{m})}}\Big].
\ee
At high temperatures, we can approximate
\be
\sum_{m} {z^{m} \over m (1 - z^{m})} = \sum_m {1 \over m^2 \beta} + \Or[1] = {\pi^2 \over 6 \beta} \left(1 + \Or[\beta] \right).
\ee
and therefore,
\be
Z(\beta) = e^{2 \pi^2 \over 6 \beta},
\ee
where we have dropped the $\Or[\beta]$ errors that should be understood and will not be displayed explicitly again. 

From the partition function above, we find that the temperature and the energy (at large energies) are related through
\be
\label{ebetarel}
E = {2 \pi^2  \over 3 \beta^2}.
\ee
Moreover, the degeneracy of states at a given energy is given by
\be
\label{fuzzentropy}
S(E) = {4 \pi^2 \over 3 \beta} = 2  \pi \sqrt{2  E \over 3} = 2 \pi \sqrt{2 \none \nfive \over 3}.
\ee

To evaluate the expression in \eqref{thermaltraceexpr}, we need to evaluate one more infinite sum over $n$.
\be
\begin{split}
& \sum_{n = 1}^{\infty} {n e^{-\beta n }  \over 1 - e^{-\beta n}}  = \sum_{n=1,m=1}^{\infty} n e^{-\beta n m} =  \sum_{m=1}^{\infty} {e^{-\beta m  } \over (1 - e^{- \beta m } )^2}   \underset{\beta \rightarrow \infty}{\longrightarrow} {\pi^2 \over 6 \beta^2}+ \Or[{1 \over \beta}].
\end{split}
\ee

Therefore we find at ``high temperatures'' that
\be
\begin{split}
\expgenfnbeta[\chi, \alpha] &= \int \prod_k d g_k  d t {d s \over L}  \left({t \over \pi}\right)^{2} e^{\sum_k\Big[-\big(\frac{\pi ^2 \mu ^2 t^2}{3 \beta } + t \big) g_k^2 -g_k \big(\frac{i \pi ^2 \mu^2 t  \alpha _k}{3 \beta }+2 i t x_k \big)+\frac{\pi ^2 \mu ^2 \alpha _k^2}{12 \beta }+\frac{\pi ^4 \mu ^2 \chi _k^2}{3 \beta ^2 L^2} \Big]}    \\
&= \int d t {d s \over L} \frac{9 \beta ^2 e^{\left(\frac{\pi ^4 \mu ^2 {\vec{\chi}}^2}{3 \beta ^2 L^2}+\frac{\pi ^2 {\vec{\alpha}}^2 \mu ^2+4 t \left(\pi ^2 \mu ^2 {\vec{x} \cdot \vec{\alpha}}-3 \beta  r^2\right)}{4 \left(3 \beta +\pi ^2 \mu ^2
   t\right)}\right)}}{\left(3 \beta +\pi ^2 \mu ^2 t\right)^2} \\
&= \frac{36 \beta ^2 e^{\frac{\pi ^4 \mu ^2 {\vec{\chi}}^2}{3 \beta ^2 L^2}-\frac{3 \beta  r^2}{\pi ^2 \mu ^2}} \left(e^{\frac{\pi ^2 {\vec{\alpha}}^2 \mu ^2}{12 \beta }+\frac{3 \beta  r^2}{\pi ^2 \mu
   ^2}}-e^{{\vec{x} \cdot\vec{\alpha}}}\right)}{\pi ^4 {\vec{\alpha}}^2 \mu ^4+36 \beta ^2 r^2-12 \pi ^2 \beta  \mu ^2 {\vec{x} \cdot\vec{\alpha}}},
\end{split}
\ee
where $r^2 = \sum_k x^k x^k$.

From this generating function we can immediately read off the various ``thermal'' expectation values. We find that
\be
\label{allexpecs}
\boxed{
\begin{aligned}
&\thermexp[f_5  - 1] = Q_5 \frac{1-e^{-\frac{r^2}{\tauconst }}}{r^2}; \\
&\thermexp[f_1  - 1] = Q_5  \frac{24 \tauconst ^2 \left(1-e^{-\frac{r^2}{\tauconst }}\right)}{\mu ^2 L^2 r^2}  =  Q_1  \frac{ \left(1-e^{-\frac{r^2}{\tauconst }}\right)}{r^2};  \\
&\thermexp[A_i] = 0; \\
&\thermexp[\wg_i] = -Q_5 \frac{\tauconst x_i e^{-\frac{r^2}{\tauconst }} \left(1  - e^{\frac{r^2}{\tauconst }}+{r^2 \over \tauconst} \right)}{r^4}. \\
\end{aligned}}
\ee
where 
\be
\label{gammadef}
\tauconst = {\pi^2 \mu^2  \over 3 \beta},
\ee
In the second line of \eqref{allexpecs} we noted that  
\be
\label{q1q5ratio}
{24 \tauconst^2 \over \mu^2 L^2} = {8 \pi^4 \mu^2 \over 3 \beta^2 L^2} = {Q_1 \over Q_5}.
\ee
As advertised, $\thermexp[A_i]$ vanishes in the average fuzzball metric.

\subsubsection{Analysis of one-point functions}

The expressions in \eqref{allexpecs} start deviating from the conventional expressions when $r^2 = \tauconst$. To understand this physically, we consider the radius of the $y$-circle when $r^2 = \tauconst$ in the {\em conventional} metric. We see that this radius (in the Einstein frame) is given by
\be
\rstretchsq = \left({Q_5 \over Q_1}\right)^{1 \over 4} \left({1 \over Q_1 Q_5}\right)^{1 \over 2} \tauconst R^2.
\ee
Note that
we have 
\be
\tauconst = {\mu^2 \pi^2 \over 3 \beta}  =  {\mu^2 S(E) \over 4} = {g_s^2 S(E) \over 4 R^2 V_4},
\ee
where we used the fact that $\mu^2 = {g_s^2 \over R^2 V_4}$. 
We also note that the volume of the compact manifold in the string frame is given by $\vcom = \left({Q_1 \over Q_5}\right)  V_4$.

Putting all these factors together, and using \eqref{chargebrane} and \eqref{fuzzentropy}, we find
\be
\rstretchsq = \left({Q_5 \over Q_1}\right)^{1 \over 4} \left({V_4 \over g_s^2 \none \nfive}\right)^{1 \over 2} {g_s^2 2 \pi \sqrt{2 \none \nfive \over 3}  \over 4  V_4} = \left({Q_1 \over Q_5} \right)^{1 \over 4} {\pi \over 2} \sqrt{2 \over 3} {g_s  \over \sqrt{\vcom}}.
\ee
 In these units, where string scale is set to unity, the fundamental (10 dimensional) Planck scale is simply given by $\lpl^8 = g_s^2$. Therefore we have found that
\be
\label{rstretched}
\rstretchsq =  {\pi \over 2} \sqrt{2 \over 3} \left({Q_1 \over Q_5} \right)^{1 \over 4} {\lpl^4 \over \sqrt{\vcom}}.
\ee

Now, we should work in the duality frame, where the physical volume of the compact manifold is at least $\vcom \geq 1$ in string-units. If we are not in such a frame, we should use T-duality in the compact directions to reach such a frame.  Moreover, we should work in the duality frame where the dilaton does not blow up at this point in space, and therefore we need ${Q_1 \over Q_5} = \Or[1]$. Just as above, if this constraint is not met, we can use the U-duality group to change the values of $\none, \nfive$ while keeping $\none \nfive$ constant to reach such a duality frame.  Putting these physical constraints into \eqref{rstretched}, we see that  
\be
\label{rstretchedsmall}
\rstretch \ll \Or[\lpl].
\ee

Therefore, in the average fuzzball solution, the metric starts to differ from the conventional metric when the $y$-circle has a size that is smaller than the Planck length. The solution is {\em completely unreliable} since, well before this size is reached, classical string effects become important that have not been taken into account in obtaining the solution \eqref{fuzzmetric}. Moreover, in this region, as one might expect --- and as we compute explicitly in the next subsection ---  quantum fluctuations in the geometry are as large as various classical expectation values. This implies that we should {\em not}  take \eqref{fuzzmetric} with the substitutions \eqref{allexpecs} seriously as a quantum-corrected geometry.

\subsection{Fluctuations: $\typexp[ (f_5 - 1)^2], \typexp[(f_1 - 1)^2], \typexp[\wg_i \wg_j] $}
We now compute the quantum fluctuations in the thermally averaged ensemble as an input to computing the quantumness parameters for these quantities. We do not compute these parameters for  $A_i$, since it vanishes both in the conventional geometry and the black hole.  We consider the generating function of the variance
\be
\begin{split}
\fluctgenfnbeta[\chi, \alpha, \chip, \alphap] = {1 \over Z(\beta)}  \tr \Big[ e^{-\beta H} \int_{-\infty}^{\infty} \prod_k d g_k d {\gp}_k \int_0^{\infty} &d t  d \tp \int_0^L {d s d \sp \over L^2} \left({t \tp \over \pi^2}\right)^{2}  :e^{\exponfn}: :e^{\exponfnpr}:  \Big],
\end{split}
\ee
where
\be
\exponfnpr = -\sum_k \tp \gp_k \gp_k  + 2 i \tp (x^k - F^k(\sp)) \gp_k +  \sum_j \alphap^j F^j(\sp) + \chip^j {d F^j(\sp) \over d \sp}.
\ee
(Note that $\exponfn$ and $\exponfnpr$ share the same value of $x$.)
We can then obtain the required two-point functions by differentiation
\be
\begin{split}
&\thermexp[(f_5-1)^2] = 
Q_5^2 \fluctgenfnbeta[\chi = 0, \chip = 0, \alpha = 0, \alphap = 0]; \\
&\thermexp[(f_1-1)^2] =  
Q_5^2 \lim_{\chi^i, \chip^i \rightarrow 0}\sum_{i, j} {\partial^4 \over \partial^2 \chi^i \partial^2 \chip^i} \fluctgenfnbeta[\chi, \chip, \alpha = 0, \alphap = 0];  \\
&\thermexp[ \wg_i \wg_j]  = Q_5^2 \lim_{\alpha^i, \alphap^i \rightarrow 0} {\partial^2 \over \partial \alpha_i \partial \alphap_j} \fluctgenfnbeta[\chi = 0, \alpha, \chip = 0, \alphap]; \\
\end{split}
\ee 
By the equivalence of ensembles these variances coincide with microcanonical variances up to terms suppressed by the entropy. Then, by result \ref{microtyp} these variances also coincide with variances computed in a typical state: 
\be
\thermexp[(f_5-1)^2] = \typexp[(f_5-1)^2] + \Or[{1 \over \sqrt{S(E)}}],
\ee
and similarly for $\thermexp[(f_1-1)^2]$ and $\thermexp[\wg_i \wg_j]$.

The computation of the variance is considerably more involved than the computation of the expectation value. So, we will not compute the function 
${\cal V}_{\beta}$ for arbitrary values of its parameters, but simply compute the derivatives and limits that we are interested in.

First we note that
\be
:e^{\exponfn}: : e^{\exponfnpr}: = :e^{\exponfn + \exponfnpr}: e^{\normconst},
\ee
where the normal ordering constant $\normconst$ arises because we need to move the creation operators inside $\exponfnpr$ past the annihilation operators of $\exponfn$. This can be easily done through the formula
\be
e^{c_n^k a_n^k} e^{d_n^k (a_n^k)^{\dagger}} = e^{d_n^k (a_n^k)^{\dagger}} e^{c_n^k a_n^k}  e^{c_n^k d_n^k}.
\ee

In our case, we find that
\be
\begin{split}
&{\cal N} = \sum_{n, k}  e^{2 \pi i n (\sp - s) \over L} \Bigg[ {1 \over n} \Big(-2 \mu ^2 t {\tp} g_k {\gp}_k-i \mu ^2 t g_k {\alphap}_k-i \mu ^2 {\tp} {\gp}_k \alpha _k+\frac{1}{2} \mu ^2 \alpha _k {\alphap}_k \Big)  \\ 
&+ \Big( \frac{2 \pi  \mu ^2 t g_k {\chip}_k}{L}-\frac{2 \pi  \mu ^2 {\tp} {\gp}_k \chi _k}{L}+\frac{i \pi  \mu ^2 \alpha _k {\chip}_k}{L}-\frac{i \pi  \mu ^2 {\alphap}_k \chi _k}{L}\Big) + \frac{2 \pi ^2 n \mu ^2 \chi _k {\chip}_k}{L^2} \Bigg] \\
&= \Bigg[-\log(1 - e^{2 \pi i  (\sp - s) \over L}) \Big(-2 \mu ^2 t {\tp} g_k {\gp}_k-i \mu ^2 t g_k {\alphap}_k-i \mu ^2 {\tp} {\gp}_k \alpha _k+\frac{1}{2} \mu ^2 \alpha _k {\alphap}_k \Big)  \\ 
&+ {1 \over (1 - e^{2 \pi i  (\sp - s) \over L})}  \Big( \frac{2 \pi  \mu ^2 t g_k {\chip}_k}{L}-\frac{2 \pi  \mu ^2 {\tp} {\gp}_k \chi _k}{L}+\frac{i \pi  \mu ^2 \alpha _k {\chip}_k}{L}-\frac{i \pi  \mu ^2 {\alphap}_k \chi _k}{L}\Big) \\ &+  {1 \over (1 - e^{2 \pi i  (\sp - s) \over L})^2} \frac{2 \pi ^2 \mu ^2 \chi _k {\chip}_k}{L^2} \Bigg]. 
\end{split}
\ee
Also,
\be
\begin{split}
\exponfn + \exponfnpr &= \sum_{n, k} {1 \over {2 L \sqrt{n}}} \\ &\times \Big[\sqrt{2} \mu  a_{n}^k \left(e^{-\frac{2 i \pi  n {\sp}}{L}} (-2 i L t g_k+L \alpha _k-2 i \pi  n \chi _k)+e^{-\frac{2 i \pi  n s}{L}} (-2 i L {\tp} {\gp}_k+L
   {\alphap}_k-2 i \pi  n {\chip}_k)\right) \\ &+\sqrt{2} \mu  (a_n^k)^{\dagger} \left(e^{\frac{2 i \pi  n s}{L}} (-2 i L t g_k+L \alpha _k+2 i \pi  n \chi _k)+e^{\frac{2 i \pi  n {\sp}}{L}} (-2 i L
   {\tp} {\gp}_k+L {\alphap}_k+2 i \pi  n {\chip}_k)\right)\\ &-2 L \sqrt{n} \left(-2 i x_k (t g_k+{\tp} {\gp}_k)+t g_k^2+{\tp} {\gp}_k^2\right)\Big].
\end{split}
\ee
Therefore,
\be
\thermexp[:e^{\exponfn + \exponfnpr}:] = e^{\sum_{n,k} {\cal Q}_{n,k}},
\ee
with
\be
\begin{split}
&{\cal Q}_{n,k} = \frac{\mu ^2 z^n}{2 n \left(z^n-1\right)}  \\ &\times \left(2  t g_k e^{\frac{2 i \pi  n {\sp}}{L}}+2 {\tp} {\gp}_k e^{\frac{2 i \pi  n s}{L}}+i  {\alphap}_k e^{\frac{2 i \pi  n s}{L}}+{2 \pi  n \over L}
   {\chip}_k e^{\frac{2 i \pi  n s}{L}}+i \alpha _k e^{\frac{2 i \pi  n {\sp}}{L}}+{2 \pi  n  \over L} \chi _k e^{\frac{2 i \pi  n {\sp}}{L}}\right) \\ 
&\times \left(2 t g_k e^{\frac{-2 i \pi  n s}{L}}+2  {\tp}
   {\gp}_k e^{\frac{-2 i \pi  n {\sp}}{L}}+i  \alpha _k e^{\frac{-2 i \pi  n s}{L}}-{2 \pi  n \over L} \chi _k e^{\frac{-2 i \pi  n s}{L}}+i  {\alphap}_k e^{\frac{-2 i \pi  n {\sp}}{L}}-{2 \pi  n \over L} {\chip}_k e^{\frac{-2 i \pi  n {\sp}}{L}}\right) \\ &-t g_k (g_k-2 i x_k)-{\tp} {\gp}_k ({\gp}_k-2 i x_k).
\end{split}
\ee

We can compute the sums over $n$ using the following formulas. As $\beta \rightarrow 0$, we have
\be
\begin{split}
&\sum_n \frac{e^{-\beta  n+2 i \pi  n s}}{n \left(1-e^{-n \beta }\right)} \rightarrow \frac{\text{Li}_2\left(e^{2 i \pi  s}\right)}{\beta }; \\
&\sum_n \frac{e^{-\beta  n+2 i \pi  n s}}{1-e^{-n \beta }} \rightarrow -\frac{\log \left(1-e^{2 i \pi  s}\right)}{\beta }; \\
&\sum_n \frac{n e^{-\beta  n+2 i \pi  n s}}{1-e^{-n \beta }} \rightarrow \frac{1}{\beta  \left(1-e^{2 i \pi  s}\right)}.
\end{split}
\ee

We can then write
\be
\label{qcoeffs}
{\cal N} + \sum_{n,k} {\cal Q}_{n,k}  = \sum_k {-(g^k, \gp^k) \cdot \begin{pmatrix} \tauconst t^2 + t & c t \tp \\ c t \tp & \tauconst \tp^2 + \tp \end{pmatrix} \cdot (g^k, \gp^k) + \ell_1^k t g^k + \ell_2^k \tp \gp^k + z},
\ee
where 
\be
\tauconst = \frac{\pi ^2 \mu ^2}{3 \beta },
\ee
as above in section \ref{subseconept} and the other coefficients take on the values
\be
\label{quantexpectcoeffs}
\begin{split}
c &= {\mu^2 \over \beta} {\cal L}^+  + c_{\cal N} ;
\\-i \ell_1^k &=   2 x_k-\frac{\mu ^2}{3 \beta} \Bigg[\pi ^2  \alpha_k+3  {\alphap}_k {\cal L}^+ -6 i \pi  {{\chip}_k \over L}{\cal L}^-\Bigg] - i \ell_{1 {\cal N}}^k ;\\
-i \ell_2^k &=  2 x_k-\frac{\mu ^2}{3 \beta} \Bigg[\pi ^2  (\alphap)_k + 3 \alpha _k{\cal L}^+ +6 i \pi {\chi _k  \over L}{\cal L}^-\Bigg]  - i \ell_{2 {\cal N}}^k; \\
z &= \frac{\mu ^2}{12 \beta ^2 } \Bigg[\pi ^2 \beta   \alpha _k^2+   \left(-24 \pi^2  \beta  {\chi_k {\chip}_k \over L^2} + 4\pi ^4 {\chi _k^2 \over L^2} +4 \pi ^4 {{\chip}_k^2 \over L^2} \right)+\pi^2  \beta   {\alphap}_k^2  \\ &+12 i \pi \beta {\alphap}_k {\chi _k \over L}{\cal L}^- +   \alpha _k  {\alphap}_k
  {\cal L}^+  -12 i \pi \beta  {{\chip}_k \over L}{\cal L}^-\Bigg] + z_{\cal N},
\end{split}
\ee
where
\be
\begin{split}
&{\cal L}^+ =  {Li}_2\left(e^{-\frac{2 i \pi  (s-{\sp})}{L}}\right)+{Li}_2\left(e^{\frac{2 i \pi  (s-{\sp})}{L}}\right), \\
&{\cal L}^{-} =  \log \left(1-e^{-\frac{2 i \pi  (s-{\sp})}{L}}\right)-\log \left(1-e^{\frac{2 i \pi  (s-{\sp})}{L}}\right).
\end{split}
\ee
The contribution from the normal ordering constant in the coefficients above is indicated with the subscript ${\cal N}$ and these numbers are given by
\be
\label{normalordcoeffs}
\begin{split}
&c_{\cal N} = \mu ^2  {\cal L}_{\cal N}; \\
&{\ell}_{1 {\cal N}} = -i \mu ^2 \big(i {\alphap}_k {\cal L}_{\cal N}+\frac{2 \pi  {\chip}_k}{L \left(-1+w\right)}\big); \\
&{\ell}_{2 {\cal N}} = -i \mu ^2\big(i \alpha _k {\cal L}_{\cal N}-\frac{2 \pi  \chi _k}{L \left(-1+w\right)}\big); \\
&z_{\cal N} = -\frac{\mu ^2}{2\left(-1+w\right)^2}\\ &\times \Bigg[\alpha _k \left(-1+w\right) \left({\alphap}_k \left(-1+w\right) {\cal L}_{\cal N}  
-2 i \pi  {{\chip}_k \over L} \right) -2 \pi  {\chi _k \over L} \left(2 \pi  {{\chip}_k \over L} w-i  {\alphap}_k \left(-1+w\right)\right)\Bigg],
\end{split}
\ee
with $w = e^{\frac{2 i \pi  (s-{\sp})}{L}}$ and ${\cal L}_{\cal N} = \log(1 - 1/w)$. By comparing \eqref{normalordcoeffs} to \eqref{quantexpectcoeffs} we see that the coefficients that come from normal ordering are all negligible when $\beta \ll 1$. So, for numerical purposes we will neglect these coefficients although the fact that they are non-zero will play a  role below.

In this form we can immediately do the integral over $g_k$ and ${\gp}_k$, since these are Gaussian. We find that
\be
\fluctgenfnbeta[\chi, \alpha, \chip, \alphap] =   \int_0^{\infty} d t  d \tp \int_0^L {d s d \sp \over L^2}  \frac{e^{\frac{t {\tp} \left(4 \tauconst^2 z+\tauconst \left({\ell_1}^2+{\ell_2}^2\right)-2 c (2 c z+{\ell_1} \cdot {\ell_2})\right)+ \left(4 \tauconst z (t+{\tp})+{\ell_1}^2 t+{\ell_2}^2 {\tp}\right)+4  z}{4 \left(t
   {\tp} \left(\tauconst^2-c^2\right)+\tauconst (t+{\tp})+1\right)}}}{\left(t {\tp} \left(\tauconst^2-c^2\right)+\tauconst (t+{\tp})+1\right)^2}.
\ee

At this stage, rather than do the $t$-integrals in generality, it is convenient to separate the computation of the different quantities.

\paragraph{$\boldsymbol{\thermexp[(f_5 - 1)^2]}$\\}
This is the simplest $t$-integral to calculate since we simply set $\chi, \chip, \alpha, \alpha$ to $0$.  With these substitutions, $z \rightarrow 0$ and also $\ell_1^k, \ell_2^k \rightarrow 2 i x^k$.  We then find that
\be
\label{f5sqresult}
\begin{split}
&\fluctgenfnbeta[\chi = 0, \alpha = 0, \chip =0, \alphap =0] = \int_0^L {d s \over L} {d \sp \over L} \int_0^{\infty} d t d \tp 
\frac{\exp \left(-\frac{r^2 (2 t \tp (\tauconst-c)+ (t+\tp))}{t \tp \left(\tauconst^2-c^2\right)+\tauconst  (t+\tp)+1}\right)}{\left(t \tp \left(\tauconst^2-c^2\right)+\tauconst (t+\tp)+1\right)^2} \\
&=\int_0^L {d s  \over L} {d \sp \over L} \Big[{e^{-{r^2 \over c}} \over c^2} \bigg({\text{Ei}\left(\frac{r^2}{c}\right) -2 \text{Ei}\left(\frac{(\tauconst-c) r^2}{\tauconst c}\right)+\text{Ei}\left(\frac{(\tauconst-c) r^2}{c (\tauconst+c)}\right)} \bigg) 
\\ &\hphantom{\int_0^L {d s  \over L} {d \sp \over L}}
+ \frac{2 \tauconst e^{-\frac{r^2}{\tauconst}}}{c r^2 (\tauconst-c)}-\frac{(\tauconst + c) e^{-\frac{2 r^2}{\tauconst+c}}}{ c r^2 (\tauconst-c)}-\frac{1}{c r^2 } \Big].
\end{split}
\ee
Here $Ei(x) = -\int_{-x}^{\infty} e^{-t} {\cal P}\left({1 \over t}\right) d t$
The last integral, over $s, \sp$ must be done numerically and we discuss that in section \ref{subsecdiffquantf1f5}.

\paragraph{$\boldsymbol{\thermexp[(f_1- 1)^2]}$\\}
It is convenient to first differentiate with respect to $\chi, \chip$ before performing the $t, \tp$ integrals. These derivatives lead to a complicated expression. However, it is important to realize that this expression involves terms that appear at different orders in ${1 \over \beta}$. 

The reader can persuade herself through inspection, or through an explicit calculation that the dominant terms at small $\beta$ occur only when the $\chi, \chip$ derivatives act on the $\chi_k^2$ and $(\chip^k)^2$ terms inside $z$ in \eqref{quantexpectcoeffs}. 

This leads to the simple result 
\be
\label{f1sqresult}
\thermexp[(f_1 - 1)^2]  = {64 \pi^8 \mu^4 \over 9 L^4 \beta^4} \thermexp[ (f_5 - 1)^2] \left(1  + \Or[\beta] \right) = {Q_1^2 \over Q_5^2} \thermexp[ (f_5 - 1)^2] \left(1  + \Or[\beta] \right),
\ee
where we have used the ratio between coefficients displayed in \eqref{q1q5ratio} and neglected the $\Or[\beta]$ terms in the last equality.

\paragraph{${\boldsymbol{\thermexp[\wg_i \wg_j]}}$  \\} 
The computation of $\thermexp[ \wg_i \wg_j]$ is rather involved. We do not give the details of all the intermediate steps, but simply note the final answer in the form of an integral over $s, \sp$ that we will evaluate numerically below. We find that
\be
\label{aisqresult}
{1 \over Q_5^2} \thermexp[\wg_i \wg_j] =  {\cal A} \delta_{i j} + {\cal B} x_i x_j,
\ee
where
\be
\begin{split}
{\cal A} =& \frac{(c-\tauconst) e^{-\frac{r^2}{c}} \left(r^2 (c-\tauconst)+3 c (\tauconst+c)\right) \left(-2 \text{Ei}\left(\frac{(\tauconst-c) r^2}{\tauconst c}\right)+\text{Ei}\left(\frac{(\tauconst-c) r^2}{c (\tauconst+c)}\right)+\text{Ei}\left(\frac{r^2}{c}\right)\right)}{12 c^4} \\
&+ \frac{\tauconst e^{-\frac{r^2}{\tauconst}} \left(-\tauconst c^2 (\tauconst+3 c)+r^4 (\tauconst-c)^2+c r^2 (c-\tauconst) (2 \tauconst+3 c)\right)}{6 c^3 r^4 (\tauconst-c)} \\
&+\frac{(\tauconst+c) e^{-\frac{2 r^2}{\tauconst+c}} \left(c^2 r^2 (\tauconst+c)^2+r^6 \left(-(\tauconst-c)^2\right)+2 c r^4 (\tauconst-c) (\tauconst+c)\right)}{12 c^3 r^6 (\tauconst-c)},
\end{split}
\ee
and
\be
\label{aijresult}
\begin{split}
{\cal B} =& \frac{\left(\tauconst^2+4 \tauconst c+c^2\right) e^{-\frac{r^2}{c}} \left(-2 \text{Ei}\left(\frac{(\tauconst-c) r^2}{\tauconst c}\right)+\text{Ei}\left(\frac{(\tauconst-c) r^2}{c (\tauconst+c)}\right)+\text{Ei}\left(\frac{r^2}{c}\right)\right)}{6 c^4} \\
&+ \frac{\tauconst e^{-\frac{r^2}{\tauconst}} \left(\tauconst^2 \left(2 c^2+c r^2+r^4\right)+\tauconst c \left(6 c^2+5 c r^2+4 r^4\right)+c^2 r^2 \left(6 c+r^2\right)\right)}{3 c^3 r^6 (\tauconst-c)} \\
&-\frac{(\tauconst+c) e^{-\frac{2 r^2}{\tauconst+c}} \left(\tauconst^2 \left(2 c^2+c r^2+r^4\right)+2 \tauconst c \left(2 c^2+3 c r^2+2 r^4\right)+c^2 \left(2 c^2+5 c r^2+r^4\right)\right)}{6 c^3 r^6 (\tauconst-c)}.
\end{split}
\ee

\subsection{Analysis of the results}
We now proceed to analyze the results obtained. We will compute the ``difference'' and ``quantumness'' parameters defined in \eqref{diffparamdef} and \eqref{quantparamdef}. For the harmonic functions,  we compute the following expressions both of which depend on $\vec{x}$.
\be
\label{f5diffclassparams}
\begin{split}
&\devpar_5 = \Big|{\thermexp[(f_5 - 1)]  - f_5^{\text{bh}} + 1\over \thermexp[(f_5 - 1)]}\Big|; \\
&\quantpar_5 = \Big|{\left(\thermexp[(f_5 - 1)^2]  - \thermexp[(f_5 - 1)]^2 \right)^{1 \over 2} \over \thermexp[ (f_5 - 1)]} \Big|. 
\end{split}
\ee
We do not need to compute these parameters separately for $f_1$ since as we have found above, $\thermexp[f_1 - 1] = {Q_1 \over Q_5} \thermexp[f_5 - 1]$ and also $\thermexp[(f_1 - 1)^2] = {Q_1^2 \over Q_5^2} \thermexp[(f_5 - 1)^2]$. Therefore $\quantpar_1 = \quantpar_5; ~\devpar_1 = \devpar_5$.

Since $\wg_i = 0$ in the conventional black-hole geometry, we find $\devpar_{\wg} = 1$. Then, to measure the size of quantum fluctuations, we compute
\be
\quantpar_{\wg} = \Big| {\left({{x}^i {x}^j \thermexp[ \wg_i \wg_j ] - {x}^i {x}^j \thermexp[ \wg_i ] \thermexp[ \wg_j ]}\right)^{1 \over 2} \over   {x}^i {x}^j \thermexp[ \wg_i ] \thermexp[ \wg_j ]} \Big|.
\ee

\subsubsection{\bf Difference and quantumness parameters for the harmonic functions, $f_1$ and $f_5$ \label{subsecdiffquantf1f5}}
We start by evaluating \eqref{f5diffclassparams}.  Let us consider the remaining integrals over $s, \sp$ in the results  \eqref{f5sqresult}.

First note that $s, \sp$ only appear in these expressions through  $w = e^{{2 \pi i (s - \sp) \over L}}$. Moreover, each integrand has a  Laurent series expansion in $w$, and so the integral over $s, \sp$ simply picks out the $w^0$ term. Note also, that the integrand has potential singularities at $\tauconst = c$. These poles arise when $w$ solves
\be
{\pi^2 \over 6} -  {1 \over 2} \left(\pl(2, w) + \pl(2, 1/w)  + {2 \pi \beta \over L} \log(1 - w)\right) = 0.
\ee
At small $\beta$ the pole is {\em almost} at $w = 1$  but the last term moves the pole slightly off the $|w| = 1$ contour to $|w| = 1 + \Or[\beta]$. 
Other than the effect above, the normal ordering term is negligible. For ease with the final numerical integrals, we deal with this as follows. We drop the term $c_{\cal N}$ in \eqref{quantexpectcoeffs} and instead add a small real regulator to $c$ so that $c_{\epsilon} = c - \epsilon$. We then compute \eqref{f5sqresult} and \eqref{aisqresult} with $c \rightarrow c_{\epsilon}$. Numerically, it is easy to check that the value of this regulator does not alter any of the answers provided it is kept small enough.

It is of interest to consider the behaviour near $r = 0$, where the fuzzball solution differs from the conventional solution. We find that, in this limit \eqref{f5sqresult}
\be
\begin{split}
{1 \over Q_5^2} \thermexp[ (f_5 - 1)^2 ] = &Q_5^2 \int_0^L {d s d \sp \over L^2}  \frac{\log \left(\frac{\tauconst^2}{\tauconst^2-c^2}\right)}{c^2}+\frac{r^2 \left(\tauconst (\tauconst+c) \log \left(1-\frac{c^2}{\tauconst^2}\right)+c^2\right)}{\tauconst c^3 (\tauconst+c)}+O\left(r^4\right) \\
&= 1.182 {1 \over \tauconst^2} - 1.283 {r^2 \over \tauconst^3},
\end{split}
\ee
which leads to
\be
\quantpar_5 = 0.426 - 0.119 {r^2 \over \tauconst}.
\ee

They key point is that, just as expected $\quantpar_5$ becomes of order $1$ just in the region where ${r^2 \over \tauconst} \ll 1$ and the average fuzzball geometry starts to differ appreciably from the conventional solution. Therefore, in precisely the region where the average fuzzball geometry predicts interesting effects, it becomes unreliable.

\begin{figure}[!h]
\begin{center}
\includegraphics[width=0.4\textwidth]{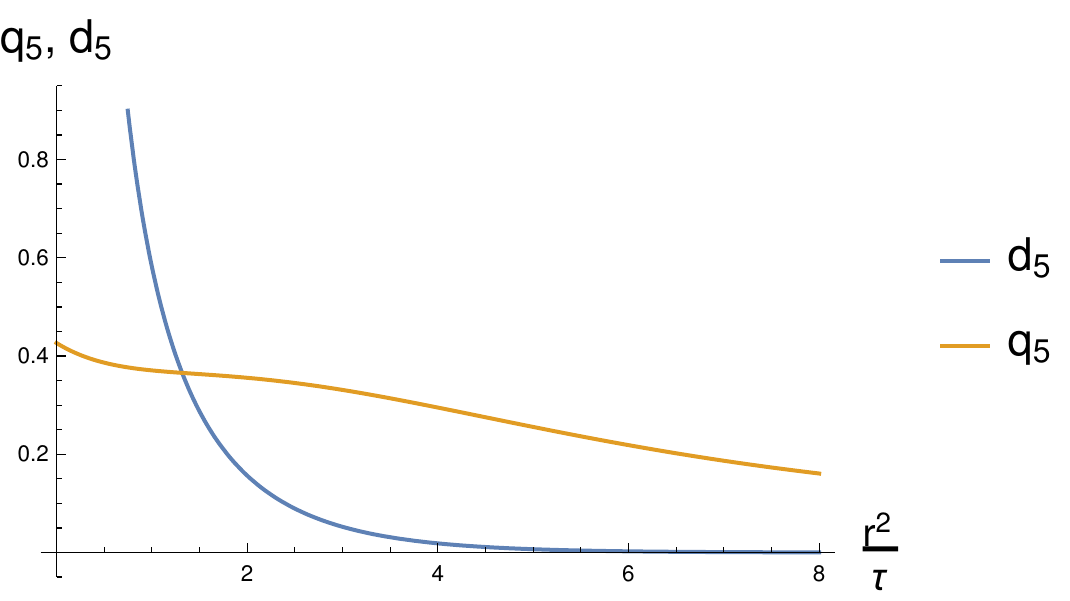}
\caption{\em A plot of the ``difference parameter'', $\devpar_5$ and the ``quantumness parameter'', $\quantpar_5$ for the harmonic functions $f_5$ and $f_1$.  The plot shows that in the average fuzzball geometry, precisely in the region where $f_5, f_1$ differ significantly from their conventional value, quantum fluctuations becomes large and the fuzzball geometry becomes unreliable.  \label{figf5diffclassparams}}
\end{center}
\end{figure}

It is possible to numerically compute both $\devpar_5$ and $\quantpar_5$ for larger values of $x$ and we display this in Figure \ref{figf5diffclassparams}. As we move to larger values of ${r^2 \over \tauconst}$ the solution becomes more reliable, but then it also becomes indistinguishable from the conventional black hole.

\subsubsection{\bf Difference and quantumness parameters for $\wg_i$}
We now turn to the difference and quantumness parameters for $\wg_i$. The difference parameter is uniformly equal to $1$ since $\wg_i = 0$ in the conventional black hole. 

At small ${r^2 \over \tauconst}$ we find that
\be
\begin{split}
&\hat{x}^i \hat{x}^j \thermexp[ \wg_i \wg_j ] \\ &=\int_0^L {d s \over L} {d \sp \over L}\Bigg[ \frac{(\tauconst^2-c^2)  \log \left(\frac{\tauconst^2-c^2}{\tauconst^2} \right)+c^2}{4 c^3}-\frac{r^2 (\tauconst+c) \left(\tauconst^2 \log \left(\frac{\tauconst^2 - c^2}{\tauconst^2}\right)+c^2\right)}{2 \tauconst c^4}\Bigg] + \Or[r^4 \over \tauconst^3] \\ =&  {0.00489 \over \tauconst} + {0.355 r^2 \over \tauconst^2} + \Or[{r^4 \over \tauconst^3}].
\end{split}
\ee
At small values of $r$, when we combine this with the series expansion of 
\eqref{allexpecs},  this leads to 
\be
\quantpar_{\wg} = 0.140 {\sqrt{\tauconst} \over r} + 1.587 {r \over \sqrt{\tauconst}}.
\ee
Therefore, quantum fluctuations diverge near $r = 0$ and the expected value of $\wg_i$ in the average geometry becomes completely unreliable.

\begin{figure}[!h]
\begin{center}
\includegraphics[width=0.4\textwidth]{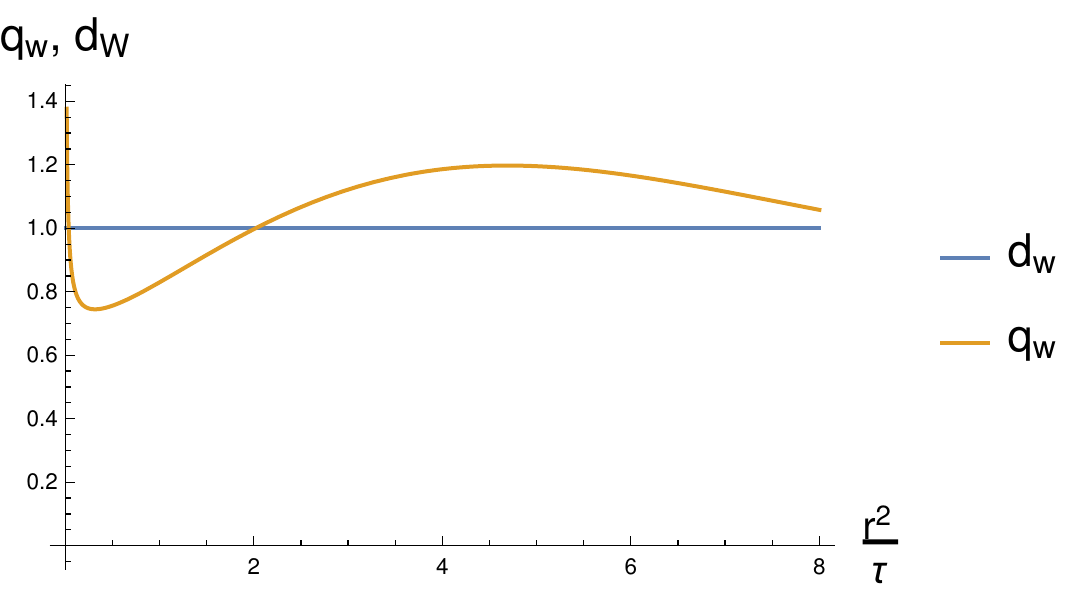}
\caption{\em A plot of the ``difference parameter'', $\devpar_{\wg}$ and the ``quantumness parameter'', $\quantpar_{\wg}$ for the one-form $\wg_i$. The difference parameter is uniformly $1$, since this function vanishes in the conventional black hole. But, the value it ostensibly takes in the average fuzzball solution is always unreliable since quantum fluctuations are the same size as its expectation value \label{figaiparams}}
\end{center}
\end{figure}
We can numerically plot the quantumness parameter for larger values of $r$ and this leads to the curve shown in Figure \ref{figaiparams}. At both small ${r^2 \over \tauconst}$ and larger values of ${r^2 \over \tauconst}$ the value for $\wg_i$ given by the average fuzzball geometry is unreliable. 

\paragraph{Fuzzballs and entropy counting.}
In this section, we have argued that the average fuzzball geometry is unreliable in the interesting region near $r = 0$. By result \ref{microtyp}, the geometry corresponding to a typical state in the Hilbert space produced by quantizing fuzzball solutions is also unreliable near $r = 0$. Away from $r = 0$, this average geometry is effectively indistinguishable from the conventional geometry since its deviation from the conventional geometry is of the same order as quantum fluctuations.

The reader might wonder how --- in spite of this fact --- counting the entropy of fuzzball solutions succeeds in getting the correct form for the number of Ramond ground states in the D1-D5 system. 

The puzzle is made more acute by recognizing that most of the contribution to this entropy comes from Planck-size fuzzballs. For small $\beta$, the reader can check that $\tauconst$ also measures the average size of the profile function 
\be
\label{tauconstsize}
\sum_i \int_0^L \thermexp[:F^i(s) F^i(s):] {d s \over L}. = 2 \tau.
\ee

Fluctuations in the size of the fuzzball are also controlled by $\tauconst$. At high temperatures,
\be
\label{sizefluctuations}
\sum_{i, j}\int_0^L \thermexp[:F^i(s)F^i(s): :F^j(\sp)F^j(\sp):] {d s \over L} {d \sp \over L} = {22 \over 5} \tauconst^2.
\ee
 This tells us that if we consider the entropy corresponding to fuzzballs that are {\em parametrically} larger than \eqref{tauconstsize}, then this entropy is highly suppressed. 

We do not have a complete explanation for the fact that quantizing the space of fuzzballs gives approximately the correct counting --- even though individual fuzzball solutions that contribute dominantly to the entropy are unreliable. Our best guess is as follows. We can consider the solutions \eqref{fuzzmetric} for profile functions, $F^i(s)$ that are {\em parametrically larger} than \eqref{tauconstsize}. In this regime, the solutions are reliable. Since they also saturate the BPS condition,  they must correspond to some ground-states of the D1-D5 system. This subclass of solutions can be quantized and counted reliably.   Then --- perhaps as a result of one of the fortuitous coincidences that occur while counting supersymmetric states --- this counting formula can be extrapolated to obtain a count of all ground states. Perhaps this last step can be explained by going to some other point in parameter space,   where these solutions can be mapped to better controlled states; this deserves to be understood better.

 However, we emphasize that the entropy-formula itself  cannot be taken as evidence that fuzzballs are giving us an accurate picture of physics near $r = 0$ at the supergravity point in moduli space,  since the solutions that dominate the entropy are unreliable in that region. 
It would be nice to understand the physics near $r = 0$ better but this clearly requires some other technique.

Another interesting open question is as follows. What is the basis whose elements minimize $\quantpar_5, \quantpar_1, \quantpar_{\wg}$ so that the fluctuations in a typical state --- as  calculated in \eqref{f5sqresult}, \eqref{f1sqresult} and \eqref{aisqresult} ---  come from differences between elements of this basis rather than fluctuations of the operator within given basis states?  We do not expect, that even in such a basis, $\quantpar_5, \quantpar_1, \quantpar_{\wg}$ will be parametrically suppressed near $r = 0$, but in such a basis the fuzzball microstates would be reliable as possible.

\section{Probing multi-charge solutions  \label{secmulticharge}}
In the previous section, we provided a detailed discussion of the two-charge fuzzballs. However, the corresponding conventional solution has vanishing horizon area. So, it is of interest to investigate fuzzballs that have the same charges as black holes with a macroscopic horizon.

In this section, we will consider the class of asymptotically AdS fuzzball solutions discovered in \cite{Bena:2016ypk} following previous work in \cite{Bena:2015bea}. The conventional black-hole geometry corresponding to these charges is given in \cite{Cvetic:1998xh}. We cannot repeat the analysis of section \ref{sectwocharge} and compute quantum expectations and fluctuations since  (a) all fuzzball solutions with the charges of \cite{Bena:2016ypk} have not been discovered and (b) these solutions have not been quantized. Moreover, even in the set of solutions of \cite{Bena:2016ypk}, we will consider only a subset for which the form of the metric was explicitly given in \cite{Bena:2016ypk}. 

These solutions are macroscopically distinguishable from the conventional black hole. Therefore, if we believe expectation \ref{euclideanaverage} then result \ref{thatypical} immediately tells us that they cannot be typical elements of a basis. The calculations in this section show that one can reach this conclusion even without assuming expectation \ref{euclideanaverage} and simply by considering asymptotic observables.

The solutions of \cite{Bena:2016ypk} are asymptotically AdS, and the asymptotic observable we will focus on is a two-point function of a marginal scalar operator in the boundary CFT. We will use this two-point function to investigate the energy-gap between successive excitations of the fuzzball solutions. We find that this gap is too large and violates expectation \ref{continuousspectrum}. We will also investigate whether fuzzball solutions satisfy a specific bound for the falloff of the two-point function for large spacelike momenta. This bound holds in any conformal field theory, and is saturated by the black-hole geometry. However, fuzzballs fail to saturate this bound, indicating that they cannot be typical microstates.

\subsection{Review of the solution}
The metric given in \cite{Bena:2016ypk} is
\be
\label{metricmulticharge}
\begin{split}
&ds_6^2 = -{2 \over \sqrt{{\cal P}}} (d v + \beta) (d u + \omega + {1 \over 2} {\cal F} (d v  + \beta)) + \sqrt{{\cal P}} d s_4^2; \\
&u = (t - y)/\sqrt{2}; v = (t + y)/\sqrt{2}; y \sim y + 2 \pi R_y; \\
&ds_4^2 = {\Sigma d r^2 \over r^2 + a^2} + \Sigma d \theta^2 + (r^2 + a^2) \sin^2 \theta d \phi^2 + r^2 \cos^2 \theta d \psi^2;\\
&{\cal P} = Z_1 Z_2 - Z_4^2;  \quad \beta = {a^2 R_y \over \sqrt{2} \Sigma} (\sin^2 \theta d \phi - \cos^2 \theta d \psi); \quad \Sigma = (r^2 + a^2 \cos^2 \theta).
\end{split}
\ee
The functions $Z_1, Z_4, {\cal F}$ in the metric above depend on three integer parameters $k, m, n$. In this paper, we will consider the simplest case, $k = 1, m = 0$, $n$ arbitrary, for which it is easy to write down explicit expressions for these quantities. For this case, we find that
\be
\begin{split}
&Z_1 = \frac{\left(a^2+r^2\right)^{-n-1} \left(a^2 b^2 {R_y}^2 \sin ^2(\theta ) r^{2 n} \cos \left(2 \left(\frac{n
   (t+y)}{{R_y}}+\phi \right)\right)+2 {Q_1} {Q_5} \left(a^2+r^2\right)^{n+1}\right)}{2 {Q_5}
   \left(a^2 \cos ^2(\theta )+r^2\right)};  \\
&Z_4 = \frac{a b {R_y} \sin (\theta ) r^n \left(a^2+r^2\right)^{-\frac{n}{2}-\frac{1}{2}} \cos \left(\frac{n
   (t+y)}{{R_y}}+\phi \right)}{a^2 \cos ^2(\theta )+r^2}; \\
&{\cal F} = -\frac{1-r^{2 n} \left(a^2+r^2\right)^{-n}}{a^2}.
\end{split}
\ee 

The parameters in the solution are related through the constraint 
\be
\label{abconstraint}
{Q_1 Q_5 \over R_y^2} = a^2 + b^2/2.
\ee
The asymptotic geometry of these solutions is $\text{AdS}_3 \times S^3$ with an AdS radius $\lambda = (Q_1 Q_5)^{1 \over 4}$. By the standard formula for the AdS central charge,  we also have 
\be
{3 \lambda \over 2 G_3} = 6 \none \nfive,
\ee
where $\none, \nfive$ are the number of D1 and D5 branes and $G_3$ is the 3-dimensional Newton's constant. In addition we set ${\cal N} = \none \nfive/(a^2 + b^2/2)$.

The charges of the solutions --- the angular momenta along the $S^3$ ($J_L$, $J_R$), the mass ($M$) and the momentum along the $y$-direction $P_y$ are given by
\be
\label{multichargeformulas}
J_L = J_R = {{\cal N} \over 2} a^2; \quad M = P_y =  {{\cal N} n \over 2 R_y} b^2.
\ee

\subsection{Physical quantities of interest}

Let us consider a marginal scalar operator in the boundary theory. We will consider the Fourier transform of its {\em Wightman function} in a state dual to the fuzzball solution
\be
\label{fuzzwightman}
\wight(\freqry, \momry) = \int \langle \fm | O(t, y) O(0, 0) | \fm \rangle e^{i \freqry t \over R_y} e^{i \momry y \over R_y} d t d y.
\ee
It will also be useful to consider the Fourier transform of the {\em commutator} which is just the difference of two Wightman functions
\be
\label{fuzzcomm}
\comm(\freqry, \momry) = \int \langle \fm | [O(t, y), O(0, 0)] | \fm \rangle e^{i \freqry t \over R_y} e^{i \momry y \over R_y} d t d y = \wight(\freqry, \momry) - \wight(-\freqry,-\momry).
\ee

This Wightman function can be computed using the standard AdS/CFT dictionary by considering the boundary limit of a bulk minimally coupled massless scalar, $\phi$, with no motion along the $S^3$ coordinates.
\be
\wight(\freqry, \momry) = \lim_{r \rightarrow \infty} r^4 \int \langle \fm | \phi(r, t, y) \phi(r, 0, 0) | \fm \rangle e^{i \freqry t \over R_y} e^{i \momry y \over R_y} d t d y.
\ee
This follows from the standard ``extrapolate'' dictionary in AdS/CFT.\footnote{In some special cases in $d = 4$ that correspond to  Coulomb branch solutions of ${\cn =4}$ SYM, there are subtleties with the standard extrapolate dictionary \cite{Skenderis:2006uy} but these subtleties are irrelevant here.}

It was shown in \cite{Bena:2017upb,Tyukov:2017uig} that, in the fuzzball background under consideration, the massless scalar equation is separable. Furthermore, $t$ and $y$ are Killing vectors for the metric, and therefore it is convenient to expand the bulk scalar field as 
\be
\label{phimodeexpand}
\phi(r, t, y)= \sum_{\freqry, \momry} a_{\freqry, \momry} R_{\freqry, \momry}(r) e^{-i \freqry t \over R_y} e^{-i \momry y \over R_y}  + \text{h.c},
\ee
where the bulk operators are normalized so that $[a_{\freqry, \momry}, a^{\dagger}_{\freqry', \momry'}]  = \delta_{\freqry \freqry'} \delta_{\momry \momry'}$.
We take the radial wave-functions corresponding to different $\freqry, \momry$ to be orthonormal in the Klein Gordon norm. Therefore they satisfy
\be
\label{kleingordonnorm}
\int h(r) \freqry R_{\freqry, \momry} (r) R^*_{\freqry, \momry}(r) d r = 1.
\ee
where the measure factor, $h(r) = 8 \pi^3 \sqrt{-g}/(\sin \theta \cos \theta)$ depends  only on  $r$.
Note that this requires us to consider only {\em normalizable} radial bulk solutions.

The boundary Wightman function is then just given by
\be
\label{boundarybulkmomspace}
\wight(\freqry, \momry) = N_{\freqry, \momry} |{\casympt}_{\freqry, \momry}|^2,
\ee
where
\be
{\casympt}_{\freqry, \momry} = \lim_{r \rightarrow \infty} r^2 R_{\freqry, \momry}(r).
\ee
and 
\be
\langle \fm | a_{\freqry, \momry} a^{\dagger}_{\freqry', \momry'} | \fm \rangle = N_{\freqry, \momry} \delta_{\freqry, \freqry'} \delta_{\momry \momry'}.
\ee
If the fuzzball state is approximately thermal, we expect that $N_{\freqry, \momry} = {1 \over 1 - e^{-\beta \freqry}}$ and {\em independent} of $\momry$. However, we can avoid any assumptions about the function $N_{\freqry, \momry}$ by considering the {\em commutator}. 
Then by using the fact that 
\be
\langle \fm | [a_{\freqry, \momry}, a^{\dagger}_{\freqry', \momry'}] | \fm \rangle = \delta_{\freqry, \freqry'} \delta_{\momry \momry'}.
\ee
we see that the commutator is simply given by
\be
\comm(\freqry, \momry) =  |{\casympt}_{\freqry, \momry}|^2.
\ee

Therefore, the computation of the Wightman function and the commutator reduces essentially to a computation of ${\casympt}_{\freqry, \momry}$ which can be obtained by solving the bulk radial equation and normalizing it under \eqref{kleingordonnorm}.

We will be particularly interested in the behaviour of these function in the limit where $\momry \gg 1$. In this limit, we can perform analytic calculations using a WKB approximation. We will show that $\wight(\freqry, \momry)$ and $\comm(\freqry, \momry)$ have support on a {\em discrete} set of frequencies with a  {\em gap} between successive excitations that scales with ${a^2 \over b^2}$.  Therefore, even if we take ${a \over b} \ll 1$ (but not suppressed by an exponent of $\none \nfive$),  we see by expectation \ref{continuousspectrum} that these states are very atypical. 

The large-$\momry$ limit of the Wightman function and the commutator  is also of interest because, by virtue of having a horizon, black holes saturate a {\em bound} on how slowly these functions can decay at large-$\momry$. We will show, again, that the fuzzball solutions do {\em not} saturate this bound in the limit where ${a^2 \over b^2 n} = \Or[1]$. Therefore, these fuzzball solutions also do {\em not} obey the ETH.

\subsection{Propagation of a massless scalar}

When the angular momentum of the mode along $S^3$ is zero, the wave-equation, $\Box \phi = 0$ yields the following equation for the radial mode
\be
\begin{split}
&R_{\freqry, \momry}''(r) + Q(r) R_{\freqry, \momry}'(r) + P(r) R_{\freqry, \momry}(r) = 0; \\
&Q(r) = \frac{a^2+3 r^2}{a^2 r+r^3}; \\
&P(r) = \frac{1}{4 a^2 r^2 \left(a^2+r^2\right)^{n+2}}
 \Big(-b^4 r^2  (\freqry-\momry )^2 \left(r^{2 n}-\left(a^2+r^2\right)^n\right) -4 a^6 \momry ^2  \left(a^2+r^2\right)^n \\ &-2 a^2 r^2 \left(b^2 
   (\freqry-\momry ) \left(r^{2 n} (\freqry-\momry ) -2 \freqry \left(a^2+r^2\right)^n\right) \right)+4 a^4 r^2 \left(a^2+r^2\right)^n \left(\freqry^2-\momry ^2\right)\Big). \\
\end{split}
\ee
With the appropriate translation of notation, this is the same as the wave-equation derived in \cite{Tyukov:2017uig}.
To put the equation in WKB form, we redefine 
\be
\label{changerpsi}
R_{\freqry, \momry}(r) = \normalpha {\psifunc(r) \over \sqrt{r (r^2 + a^2)}}.
\ee
Here $\normalpha$ is a normalization constant that we will turn to in section \ref{largegammaw} and we have suppressed the $\omega, \gamma$ dependence on $\psifunc$ and $\normalpha$. Further, changing variables to $\xi = {r \over a}$ and setting $b = a \bconst$, we can put the equation for $\psifunc$ in WKB form,
\be
\label{wkbform}
{d^2 \psifunc(\xi) \over d \xi^2} - V(\xi) \psifunc(\xi) = 0,
\ee
with
\be
\label{wkbpot}
\begin{split}
V(\xi) =\frac{1}{{4 \left(\xi
   ^2+1\right)^2}}\Big[ &\frac{4 \momry ^2 - 1}{\xi ^2}+4 \momry ^2+3 \xi ^2+\bconst ^2 \left(\bconst ^2+2\right)  (\freqry-\momry )^2 \frac{\xi ^{2 n}}{\left(\xi ^2+1\right)^{n}}\\
&-\left(\bconst ^2 (\freqry-\momry )+2 \freqry\right)^2+6 \Big].
\end{split}
\ee

\begin{figure}
\label{vgraph}
\begin{center}
\includegraphics[height=0.3\textheight]{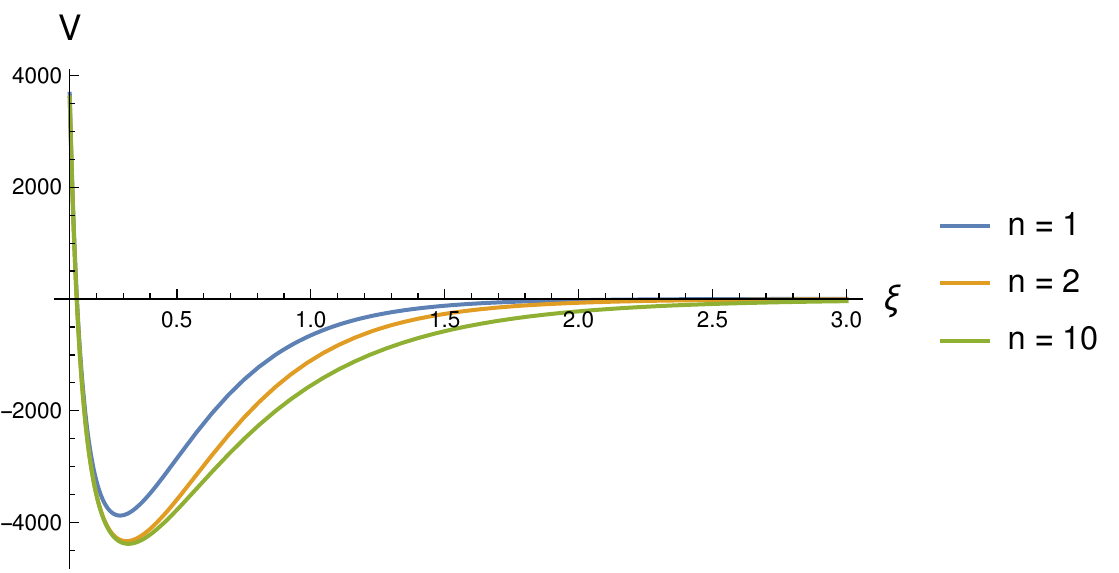}
\caption{\em A graph of $V(\xi)$ vs $\xi$ with $\momry = 10, \freqry =0, \bconst = 4$ and different values of $n$. \label{potgraph}}
\end{center}
\end{figure}

The potential has two turning points, and we can understand its qualitative behaviour as follows. We see that at small $\xi$ ($r \ll a$), the potential is positive since it is dominated by the ${4 \momry^2 - 1 \over \xi^2}$ term in the numerator of \eqref{wkbpot}. We remind the reader that $\momry \gg \freqry$ and we will primarily be interested in a regime where $\bconst \gg 1$. 
Then it is clear that for a range of values of $\xi$ near $\xi = 1$, the potential becomes negative before becoming positive again for large $\xi$ due to the $3 \xi^2$ term in the numerator. A graph of the potential is shown in Figure \ref{potgraph} for some typical values of $\bconst, n, \momry$ with $\freqry =0$ for simplicity. 

For large values of $\momry$, the WKB approximation is valid everywhere except very close to the two turning points or when $\xi = \Or[\momry]$. In this large $\xi$ region, we will match the WKB solution to a Bessel function, and we will deal with the turning points by interpolating between the two sides using the Airy-functions.

Let us denote the position of the first turning point by $\xi_1$ and the second turning point by $\xi_2$. Then, for small $\xi$ if we insist that the solution be normalizable, we can write
\be
\label{wkbpart1}
\psifunc(\xi) = {1 \over V(\xi)^{1 \over 4}}e^{\int_{\xi_1}^{\xi} \sqrt{V(\zeta)} d \zeta}, \quad \xi < \xi_1 ,
\ee
Near $\xi = \xi_1$, we can approximate $V(\xi) \approx |V'(\xi_1)| (\xi_1 - \xi)$ and therefore we have
\be
\label{wkbturn1}
\psifunc(\xi) = {2 \sqrt{\pi} \over |V'(\xi_1)|^{1 \over 6}} \Airy(|V'(\xi_1)|(\xi_1 - \xi)), \quad \xi \approx \xi_1,
\ee
where we have chosen the Airy Ai-function based on the expected asymptotics for $\xi < \xi_1$. Matching again with the WKB solution we find that
\be
\label{wkbpart2}
\begin{split}
\psifunc(\xi) =  &{1 \over |V(\xi)|^{1 \over 4}}\big[e^{i {\pi \over 4}} e^{-i \int_{\xi_1}^{\xi} \sqrt{|V(\zeta)|} d \zeta} +  e^{-i{\pi \over 4}} e^{i \int_{\xi_1}^{\xi} \sqrt{|V(\zeta)|} d \zeta} \big] \\ = &{1 \over |V(\xi)|^{1 \over 4}}\big[A_{-} e^{i \pi \over 4} e^{i \int_{\xi_2}^{\xi} \sqrt{|V(\zeta)|} d \zeta} +  e^{-i \pi \over 4} A_{+} e^{-i \int_{\xi_2}^{\xi} \sqrt{|V(\zeta)|} d \zeta} \big] , \quad \xi_1 < \xi < \xi_2,
\end{split}
\ee
where 
\be
A_{\pm} = e^{\pm i \int_{\xi_1}^{\xi_2} \sqrt{|V(\zeta)|} d \zeta}.
\ee
Near $\xi \approx \xi_2$ we need to use the Airy-function interpolation again and find that the solution matches to
\be
\label{posturntwo}
\psifunc(\xi) = {1 \over |V(\xi)^{1 \over 4}|}\big[B_{+}  e^{ \int_{\xi_2}^{\xi} \sqrt{|V(\zeta)|} d \zeta} +  B_{-} e^{-\int_{\xi_2}^{\xi} \sqrt{|V(\zeta)|} d \zeta} \big], \quad \xi_2 < \xi \ll \momry,
\ee
with $B_{+} = {A_{-} + A_{+}  \over 2}; B_{-} = {i \over 2} (A_{-} - A_{+})$. 
Now, for very large $\xi$, we can also solve for the wave-equation in terms of Bessel functions
\be
\label{largexi}
\psifunc(\xi) = {2 \casympt \over \delta} \sqrt{\xi} I_1\left({\delta \over  \xi} \right), \quad \xi \gg 1,
\ee
where $\delta^2 = {1 \over 2}(\momry^2 - \freqry^2)(2 + \bconst^2)$, and we have picked the Bessel ``I'' function by demanding that the solution be normalizable at infinity.  Matching the solutions \eqref{posturntwo} and \eqref{largexi} in the neighbourhood of some point $\xi_3$ which satisfies $1 \ll \xi_3 \ll \momry$, where both solutions are valid, we see that the solutions can only match if $
B_{+} = 0$. This simply tells us that in the region where the potential becomes positive again, the magnitude of the wave-function cannot grow exponentially.

This gives us a {\em quantization condition} on the potential: 
\be
\label{quantomega}
2 \int_{\xi_1}^{\xi_2} |V(\zeta)|^{1 \over 2} d \zeta = (2 m + 1) \pi,
\ee
for some integer $m$. From the formula above, we have $B_{-} = (-1)^m$  and this additionally tells us that 
\be
\label{asymptoticval}
{2 \casympt \over \sqrt{2 \pi}} {\xi_3 \over \delta^{3 \over 2}} e^{\delta \over \xi_3} = {(-1)^m \over V(\xi_3)^{1\over 4}} e^{-\int_{\xi_2}^{\xi_3} |V(\zeta)|^{1 \over 2} d \zeta} \implies \casympt = (-1)^m {\delta \over 2} \sqrt{2 \pi} e^{-\int_{\xi_2}^{\xi_3} |V(\zeta)|^{1 \over 2} d \zeta - {\delta \over \xi_3}}.
\ee 
It is clear that the value of $\casympt$ does not depend on the precise value of $\xi_3$ chosen to perform the matching. 

We will now use this WKB solution to compute some physical quantities of interest.

\subsection{Energy gap}
First, we consider the gap in energies of states in the neighbourhood of the fuzzball state. This is a question of the values of $\freqry$ for which $\wight(\freqry, \momry)$ has support.  

Clearly $\wight(\freqry, \momry)$ vanishes for those frequencies where no normalizable solution exists. (This is independent of any assumption about $N_{\freqry, \momry}$.) So, the energy gap can be obtained by examining the values of $\freqry$ for which the quantization condition \eqref{quantomega} is satisfied. The quantization condition can be parsed as follows. First, in the limit $\xi_1 < \xi < \xi_2$, we expand the square-root of the potential as
\be
\begin{split}
&\sqrt{|V|} = V_{{1 \over 2}, 1} \momry - \freqry \bconst^2  V_{{1 \over 2}, 0} + \Or[{1 \over \momry}],\\
&V_{{1 \over 2},1} = \frac{\sqrt{\Big|\bconst ^4 \Big(\big(\frac{1}{\xi ^2}+1\big)^n-1\Big)-2 \bconst ^2-4  \big(\frac{1}{\xi ^2}+1\big)^{n+1}\Big|}}{2 \xi^2  \left(\frac{1}{\xi ^2}+1\right)^{{n \over 2}  + 1}}, \\
&V_{{1 \over 2}, 0} =\frac{\left(\bconst ^2+2\right) \left(\left(\frac{1}{\xi ^2}+1\right)^n-1\right)\left(\frac{1}{\xi ^2}+1\right)^{-{n\over 2} - 1} }{2 \xi^2 \sqrt{\Big| \bconst ^4 \Big(\big(\frac{1}{\xi ^2}+1\big)^n-1\Big)-2 \bconst ^2-4 \big(\frac{1}{\xi ^2}+1\big)^{n+1} \Big|}}.
\end{split}
\ee
The absolute value sign inside the square-root is for later use and does not have any effect in the range under consideration.

Now, at large $\momry$, the values of $\xi_1$ and $\xi_2$  are controlled by $V_{{1 \over 2}, 1}$ and do not depend on $\freqry$. Therefore, if we consider two {\em consecutive} solutions of \eqref{quantomega} that differ by an amount $\delta \freqry$, then this difference must satisfy
\be
\label{gapanalytic}
\boxed{
\delta \freqry  = {\pi \over \bconst^2 g_n }.
}
\ee
where
\be
g_n = \int_{\xi_1}^{\xi_2} V_{{1 \over 2}, 0} d \xi.
\ee

For general values of $\bconst$, this condition can only be solved numerically but it is of interest to examine the limit where $\bconst \gg 1$. In this limit,  we can approximate $\xi_1 \approx {2 \over \bconst^2}$ and $\xi_2 \approx \bconst \sqrt{n \over 2}$. For the calculation of $g_n$, these limits are effectively $0$ and $\infty$. Therefore, expanding $V_{{1 \over 2}, 0}$ at large $\bconst$, we find
\be
g_n  =  \int_0^{\infty} d \xi \frac{\sqrt{\left(\frac{1}{\xi ^2}+1\right)^n-1} \left(\frac{1}{\xi ^2}+1\right)^{-n/2}}{2 \left(\xi ^2+1\right)} + \Or[{1 \over \bconst}].
\ee
The first few values of $g_n$ (for $n=1$ to $n=5$) are $\{0.5, 0.574,0.610, 0.632, 0.648 \}$.

Therefore, at large $\momry$ and large $\bconst$, the energy gap between consecutive excitations scales as ${1 \over \bconst^2}$ with a simple numerical prefactor. 

\subsection{Large $\momry$ Wightman function and commutator \label{largegammaw}}
We now briefly explain the significance of the behaviour of the Wightman function and the commutator at large $\momry$ but small $\freqry$. As explained in \cite{Papadodimas:2012aq}, in any conformal field theory, the large-$\momry$, small-$\freqry$, limit of the {\em thermal} Wightman function/commutator must fall off exponentially, with an exponent that is bounded below. We review this argument below. In \cite{Papadodimas:2012aq}, it was also shown that black holes saturate this bound because of the presence of the horizon. It is therefore, of interest, to understand whether fuzzballs also saturate this bound.

We will perform the analysis for the Wightman function below, although the analysis for the commutator is precisely the same. To obtain the bound on the behaviour of the Wightman function, we consider this correlator in a state with a finite temperature and chemical potential for the momentum in the $y$-direction in some arbitrary two-dimensional conformal field theory living on a circle with radius $R_y$.
\begin{equation}
\begin{split}
\wight\left(t,y\right) &= \tr\left(e^{-\beta(H - \mu P_y)} O(t, y) O(0,0) \right)  \\
&= \sum_{m, n} e^{-\beta (E_m - \mu P_m)} e^{-i t (E_n - E_m) - i y (P_n - P_m)} |\langle m | O(0,0) | n \rangle|^2,
\end{split}
\end{equation}
where the sum over $m,n$ runs over a complete set of energy/momentum eigenstates and to lighten the notation we have the same symbol for the position-space Wightman function as for its Fourier transform. Now, by Fourier transforming in time, we find that
\be
\begin{split}
&\wight(\freqry, y) = \int \wight(t, y)  e^{{i \freqry t \over R_y}} d t \\
&= 2 \pi \sum_{m, n} \delta(E_n - {\freqry \over R_y} - E_m) e^{-\beta (E_m - \mu P_m)-i (P_m - P_n) y} |\langle m| O(0,0) | n \rangle|^2 .
\end{split}
\ee
Now,  writing $y = y_r + i y_i$, and using the spectrum condition $E_m \geq |P_m|$,  we see that the real part of the exponent in the sum over $m,n$ can be written as 
\be
\begin{split}
&\text{Re}(-\beta (E_m - \mu P_m)-i (P_m - P_n) y)  \leq  -\beta E_m (1 - |\mu|)  + (|P_m| + |P_n|) |y_i|  \\
&= \beta (1 - |\mu|) {\freqry \over 2 R_y} - (E_m + E_n){\beta(1 - |\mu|)  \over 2} +  (|P_m| + |P_n|) |y_i| \\
&\leq \beta (1 - |\mu|) {\freqry \over 2 R_y} - (E_m + E_n) \left({\beta (1 - |\mu|) \over 2} - |y_i| \right).
\end{split}
\ee
Therefore the exponent always supplies a convergence factor in the sum over $m,n$ provided that $|\text{Im}(y)| < {\beta(1 - |\mu|) \over 2}$ and therefore the Green's function can be analytically continued in the $y$ plane in both directions up to this limit. But then writing
\be
\wight(\freqry, y_r - i y_i) = \int_{\momry = -\infty}^{\infty} \wight(\freqry, \momry)  e^{-i \momry (y_r + i y_i) \over R_y},
\ee
we see that, in the regime where $\momry \rightarrow \infty$, this is only possible if
\be
\lim_{\momry \rightarrow \infty} {-\log |\wight(\freqry, \momry)| \over (|\momry|/R_y)} \geq {\beta(1 - |\mu|) \over 2}.
\ee
Note that the minus sign outside the log indicates that the Wightman function must {\em decay} at large $\momry$. Second, we also note that this bound can be written in terms of the left and right temperatures that couple to the 
left and right Virasoro charges: $\beta_L = {1 \over R_y} \beta(1 - \mu); \beta_R =  {1 \over R_y}\beta(1 + \mu)$, if we recognize that $\beta(1 - |\mu|)  = R_y \text{min}(\beta_L, \beta_R)$.  In this notation the bound simply becomes
\be
\label{largegbound}
\lim_{\momry \rightarrow \infty} {-\log |\wight(\freqry, \momry)| \over \momry } \geq {1 \over 2} \text{min}(\beta_L, \beta_R).
\ee
Repeating the analysis above, we see that the {\em same} bound also applies to the commutator \eqref{fuzzcomm}
\be
\label{largegboundcomm}
\lim_{\momry \rightarrow \infty} {-\log |\comm(\freqry, \momry)| \over \momry } \geq {1 \over 2} \text{min}(\beta_L, \beta_R).
\ee

It was explained in \cite{Papadodimas:2012aq} that black holes saturate this bound. Intuitively, this happens for the following reason. In general, modes with large $\momry$ but small $\omega$ are unusual because they have larger momentum than frequency and are ``spacelike'' near the boundary. However, the black-hole horizon allows such modes to propagate in the bulk because of the red-shift near the horizon. The fuzzball also has a red-shift but we will see below that fuzzballs do {\em not} saturate \eqref{largegbound}.

To calculate the large-$\momry$ behaviour of the Wightman function and the commutator we need to compute $\normalpha$ defined in \eqref{changerpsi} and the asymptotic behaviour of $\psifunc$. We can compute
\be
\normalpha^{-2} = \int_0^{\infty} \freqry {|\psifunc(r)|^2 \over r (r^2 + a^2)} h(r) d r,
\ee
where the measure factor, $h(r)$,  is given below \eqref{kleingordonnorm}

However, we see that the WKB wave-function given in \eqref{wkbpart1},  \eqref{wkbpart2} and \eqref{largexi} has no growing exponential of $\momry$ and therefore
\be
\lim_{\momry \rightarrow \infty} {\log(\normalpha) \over \momry} = 0.
\ee

This leaves us with the asymptotic part of the wave-function, which is controlled by the coefficient ${\casympt}$ in \eqref{asymptoticval}. A simple calculation yields that at large $\momry$ we have
\be
-\log({\casympt}) = \int_{\xi_2}^{\xi_3} \sqrt{V(\xi)} d \xi + {\delta \over \xi_3} + \Or[1] =  \frac{\pi  \momry  \left(\left(8 \bconst ^2+11\right) n-1\right)}{32 \bconst ^2 n^{3/2}} + \Or[1].
\ee

Using the formula \eqref{boundarybulkmomspace} the falloff of the Wightman function  and the commutator for the fuzzball geometry is given by
\be
\label{fuzzdecay}
\lambda_{\text{fuzz}} \equiv \lim_{\momry \rightarrow \infty} {\log |\wight(\freqry, \momry)| \over \momry}  =  \lim_{\momry \rightarrow \infty} {\log |\comm(\freqry, \momry)| \over \momry} = {\pi \over 2 \sqrt{n}} + {(11 n - 1) \pi \over 16 n^{3 \over 2} \bconst^2}.
\ee

One subtlety in comparing the fuzzball result with the bound is that the fuzzballs also have angular momentum along the $S^3$. It is understood holographically that --- at least for the purpose of computing correlation functions such as \eqref{fuzzwightman}, which do not themselves depend on any $S^3$ variable ---  black holes with angular momentum along the $S^3$ direction behave as if they have  an ``effective'' Virasoro charges given by-
\be
L^{\text{eff}}_0 = L_0 - {J_L^2 \over \none \nfive},
\ee
where we have recalled that the central charge of the theory is $6 \none \nfive$. (See, for example, the discussion below (5.17) of \cite{Aharony:1999ti} and the original discussion in \cite{Breckenridge:1996sn,Breckenridge:1996is}.) We emphasize that using this ``effective charge'' rather than the original charge only {\em weakens} the bound \eqref{largegbound} and so makes the comparison more favourable for fuzzballs. 

The right inverse-temperature of the fuzzball is infinity because the solution satisfies $\bar{L}_0 = 0$. The effective left inverse-temperature corresponding to the effective charge above is given by
\be
\beta_L = \pi  \left(\bconst ^2+2\right) \sqrt{\frac{1}{\bconst ^2 \left(\bconst^2+2\right) n-1}}.
\ee
Comparing the decay of the fuzzball Wightman function and the commutator to the bound we find that
\be
\boxed{\lambda_{\text{fuzz}} - {1 \over 2} \beta_L   =  \frac{\pi  (3 n - 1)}{16 \bconst ^2 n^{3/2}}+  \Or[{1 \over \bconst^4}].}
\ee
Therefore, fuzzballs {\em fail} to saturate the large-$\momry$ bound \eqref{largegbound} by the amount shown above.

\subsection{Numerical verification}
We can verify the analytic results above by direct numerical analysis of the propagation of a scalar field in the fuzzball background. 

We consider a fuzzball background with a given value of $\bconst$ and $n$ and a scalar field excitation with a given value of $y$-momentum, $\momry$. The equation \eqref{wkbform} is subject to normalizability under the Klein-Gordon norm, and this fixes boundary conditions both at $\xi = 0$ and at $\xi = \infty$.  In fact, both $\xi = 0$ and $\xi = \infty$s are singular points of the equation, and to solve the equation numerically, we must expand in a series solution about the point $\xi = 0$ out to $\xi = \epsilon$. 

Near $\xi = 0$, we set the function and its derivative through the expansion
\be
\psifunc^{0}(\xi) = \left({\xi \over \epsilon}\right)^{2 \momry - 1 \over 2} \left(1 + a_0 \xi^2 \right), \quad 0 < \xi \leq \epsilon,
\ee
with 
\be
a_0 = -\frac{\momry ^2 \left(\bconst ^4+4\right)+\left(\bconst ^2+2\right)^2 w^2-2 \momry  \bconst ^2 \left(\bconst^2+2\right) w-8}{16 (\momry +1)}.
\ee
Note that, for numerical convenience, the normalization used here is {\em different} from the normalization used in \eqref{wkbpart1}. 

Near $\xi = \infty$, we set the function and its derivative through the expansion
\be
\psifunc^{\infty} = {1 \over \sqrt{\xi}} \left(1 + {a_{\infty} \over \xi^2} \right), \quad {1 \over \epsilon} \leq \xi < \infty,
\ee
with
\be
a_{\infty} = \frac{1}{32} \left(-\momry ^2 \bconst ^4+4 \momry ^2- \freqry^2 \bconst ^4 -4 \bconst ^2 \freqry^2-4 \freqry^2+2 \momry 
   \bconst ^4 \freqry+4 \momry  \bconst ^2 \freqry\right).
\ee
Note that, for numerical convenience, this normalization is also chosen to be different from the normalization used in \eqref{largexi}.

The allowed values of $\freqry$ can then be fixed by a {\em shooting} procedure. Given a guess for $\freqry$,  starting from $\xi = \epsilon$, we solve the equation to the mid-point of the trough of the potential: $\ximid = {1 \over 2} (\xi_1 + \xi_2)$. This solution yields some values for the function and its derivative: $\psifunc^{0}(\ximid)$ and ${d \psifunc^{0}(\ximid) \over d \xi}$. Similarly, for the same value of $\freqry$, we can start from $\xi = {1 \over \epsilon}$ and solve inwards to obtain a second set of values for the function and its derivative: $\psifunc^{\infty}$ and  ${d \psifunc^{\infty} \over d \xi}$. These values define a difference function for any given value of $\freqry$
\be
{\cal D}(\freqry) = {d \psifunc^{\infty} \over d \xi} \psifunc^{0}(\ximid) - {d \psifunc^{0}(\ximid) \over d \xi} \psifunc^{\infty}.
\ee

We then use non-linear root-finding techniques to find the roots of ${\cal D}(\freqry)$. In our analysis, we first bracketed the root, and then used the Brent method as implemented in the GNU Scientific Library \cite{galassi2009gnu}. Bracketing methods are robust and guaranteed to converge to a root in the bracketed interval. 

Note that it is because the equation is linear that we can get away by just matching the {\em ratio} of the function and its derivative at the point $\ximid$. If this had not been the case, we would have had to match both these quantities separately; this would have forced us to use two-dimensional root finding, which is far less robust. 

The asymptotic value of the function, $\casympt$, as defined above, can be fixed as follows. We denote the value that the solution starting at $\xi = 0$ takes at $\xi_1$ by $\psifunc^{0}(\xi_1)$. Then $\casympt$ is given by
\be
\casympt = {\psifunc^{0}(\ximid) \over \psifunc^{\inf}(\ximid)} \times {2 \sqrt{\pi} \Airy(0) \over |V'(\xi_1)|^{1 \over 6}} \times {1 \over \psifunc^{0}(\xi_1)}.
\ee
Note that $\Airy(0) \approx 0.355028$. 
This formula implements the following procedure: First we normalize the solution on the left so that it takes on the value given by \eqref{wkbturn1} at $\xi_1$. Then, we normalize the solution on the right so that it matches the left solution at the mid-point.

Unlike the WKB analysis, the numerical analysis is not restricted to large $\momry$. However, we can use it in the same regime to verify the results of the WKB approximation above. 

\begin{figure}[!h]
\begin{center}
\includegraphics[width=0.5\textwidth]{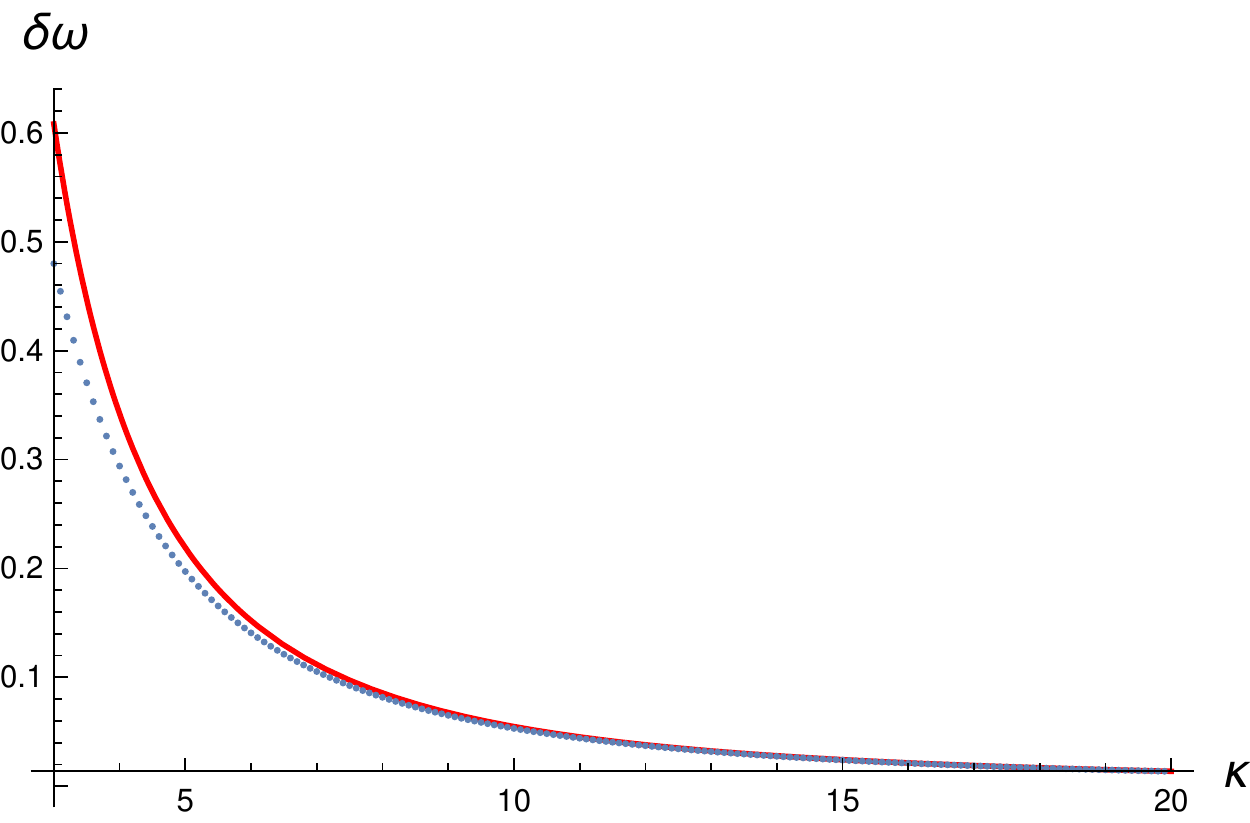}
\caption{\em Comparison between a numerical calculation (dots) of the gap between the first two allowed frequencies and the formula \eqref{gapanalytic} (solid curve). Other parameters are $\momry = 100, n=2$. In its regime of validity (large $\bconst$) the formula \eqref{gapanalytic} shows excellent agreement with the numerical results. \label{figgap}}
\end{center}
\end{figure}

\begin{figure}[!h]
\begin{center}
\includegraphics[width=0.5\textwidth]{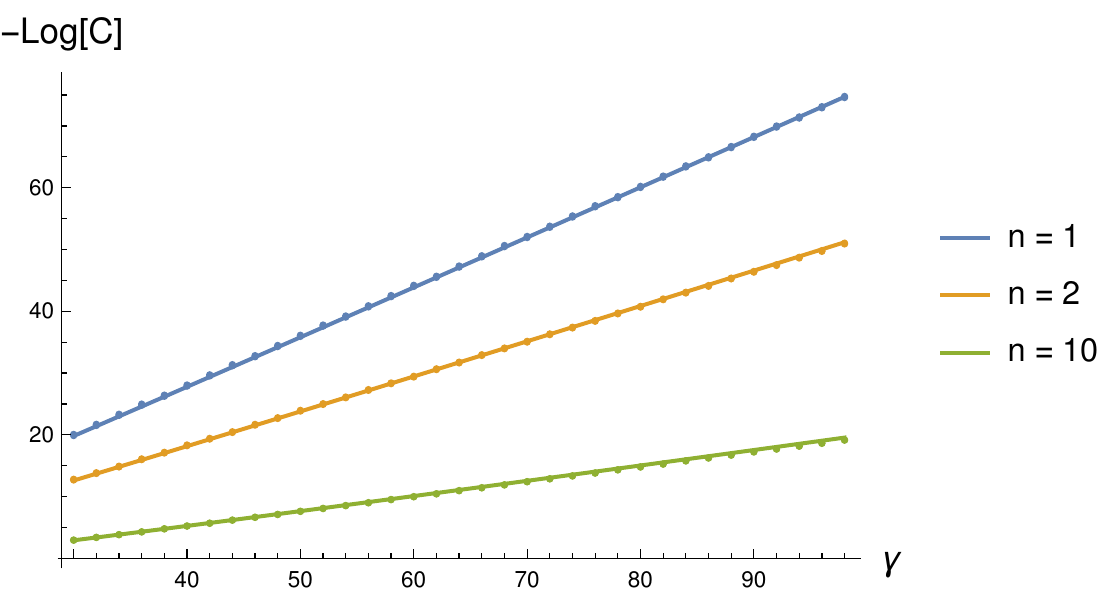}
\caption{\em Comparison between a numerical calculation (dots) of the asymptotic value $\casympt$ with the formula \eqref{asymptoticval} (solid curves) for different values of $n$ and $\momry$. Here, $\bconst = 5$ is held fixed.  In the appropriate regime of validity of the analytic formula (large $\momry$), it agrees very well with the numerical results. \label{figasymptotic}}
\end{center}
\end{figure}

In Figure \ref{figgap}, we show how the gap between the first two non-zero solutions of $\freqry$ matches with the analytic formula \eqref{gapanalytic}. 
We see that, at large $\bconst$ (which is the regime in which \eqref{gapanalytic} is derived), the agreement is excellent. In Figure \ref{figasymptotic}, we show a comparison of the numerically computed asymptotic value for $\casympt$  with the analytic formula \eqref{asymptoticval} for a fixed value of $\bconst = 5, n = 2$ and varying values of $\momry$. We see, once again, that the agreement is excellent in the regime where the analytic formula is valid.

\subsection{Analysis of the result}
The key results derived above are the formula for the mass-gap \eqref{gapanalytic} and the decay of the Wightman function \eqref{fuzzdecay} at large $\momry$, which we also verified numerically. We discuss their significance in turn.

The formula for the gap \eqref{gapanalytic} tells us that the gap between the frequency of successive excitations of these solutions is {\em far too large} for a typical microstate. As described in the discussion around \eqref{smearedgreen},  expectation \eqref{continuousspectrum} tells us  that the Wightman function (after a small amount of local smearing) should have continuous support in frequency space. As explained there, this should be true even of the Wightman function in a BPS state.  Even though the original state is BPS, the scalar excitation should connect these states to nearby non-BPS states that are expected to satisfy expectation \ref{continuousspectrum}.

However, for the fuzzball solutions we find that the support is concentrated on discretely spaced frequencies with a gap between consecutive frequencies that scales with ${1 \over \bconst^2}$. This means that the fuzzball solutions that we have analyzed, with any finite value of $\bconst$ (provided $\bconst$ does not scale with the central charge) {\em cannot} serve as microstates of a black hole. 

Instead, the finite energy gap is reminiscent of a phase of zero-entropy --- like thermal AdS as explained below \eqref{completebasisgf}. This suggests that the states corresponding to this set of fuzzball solutions belong to such a phase --- which comprises exponentially atypical states.

The formula for the decay at large-$\momry$ \eqref{fuzzdecay}  tells us that if fuzzballs are microstates, then the set of fuzzball solutions correspond to states that violate the ETH. This is because, on the basis of the black-hole calculation, the thermal state is expected to saturate the bound \eqref{largegbound}. The ETH would then suggest that eigenstates of the Hamiltonian, or indeed, typical elements of any other basis that spans that microcanonical ensemble should also saturate the bound. 

We expect that the holographic theory dual to black holes should be chaotic in supergravity regime, and therefore it should satisfy the ETH: the  hypothesis that fuzzballs are black-hole microstates contradicts this expectation. 

Second, note that even if we disregard the ETH, the idea that fuzzballs are black-hole microstates leads to a strange conclusion: since the set of fuzzballs we have analyzed are below the bound, there must be some other fuzzballs that violate  the bound \eqref{largegbound}. This is the only way that the microcanonical average can saturate the bound. Now, strictly speaking, this is not a contradiction since the bound \eqref{largegbound} is a bound for the behavior of the thermal state in different {\em theories} and does not control the behaviour of specific pure states. But, on the other hand, we are not aware of {\em any} geometry that violates the bound \eqref{largegbound}.   In the absence of such an example, it seems difficult to understand how the fuzzball geometries could represent black-hole microstates.

\paragraph{Very large values of $\bconst$.}  
We now briefly consider the limit where we take $a \sim \lpl$. In this limit $\bconst$ becomes large, and the decay of the large-$\momry$ Wightman function approaches the bound  \eqref{largegbound}. The energy-gap becomes small.\footnote{This is the gap computed in \cite{Tyukov:2017uig} by taking $\bconst = \none \nfive$.} However, this energy-gap is still {\em too large} since it scales with an inverse power of the central charge rather than being exponentially suppressed. So, even in this limit, the fuzzball solutions do not yield the correct gap expected in the boundary theory. This, by itself, ensures that even Planck-sized fuzzballs do not have the right properties expected of a typical microstate.

We now discuss some independent problems with the idea of considering fuzzballs with Planckian features. We will argue that for $a \sim \lpl$, such solutions become indistinguishable from the black hole in most of space, and quantum fluctuations are likely to be large in the near-horizon region where the fuzzball deviates from the black hole.

First, note that in this limit, the angular momentum in the $S^3$ directions, which is proportional to $a^2$, vanishes. So, in the subsector under consideration, solutions with very long throats (small values of $a$) cannot correspond to states with arbitrary charges.  However, for the remainder of this section, we will assume that when the full set of fuzzball-geometries is found, it will be possible to keep $a$ arbitrarily small, while keeping the charges constant by changing some other parameters.

Second, we note that the $a \rightarrow 0$ limit of \eqref{metricmulticharge} does not commute with the $r \rightarrow 0$ limit. 
If we first take $a \rightarrow 0$  so that we can neglect terms of order ${a \over r}$, the metric has the following smooth limit:
\be
\begin{split}
ds_6^2 \underset{a \rightarrow 0}{\longrightarrow} &\frac{ \left(b^2 n-2 r^2\right)}{\sqrt{2} b {R_y}} {dt}^2 +\frac{\left(b^2 n+2 r^2\right)}{\sqrt{2} b {R_y}} {dy}^2  +\frac{b  {R_y}}{\sqrt{2} r^2} {dr}^2 +\frac{\sqrt{2} b  n}{{R_y}} {dt} {dy}\\&+\frac{b 
   {R_y} \cos ^2(\theta )}{\sqrt{2}} {d\psi }^2 +\frac{b  {R_y} \sin ^2(\theta )}{\sqrt{2}} {d\phi }^2 + \frac{b  {R_y}}{\sqrt{2}} {d\theta
   }^2.
\end{split}
\ee
A change of variables to $\rho = \left(r^2 + {b^2 n \over 2} \right)^{1 \over 2}$ shows that this is the metric of an extremal BTZ black hole $\times$ $S^3$.

On the other hand, even when ${b \over a} \gg 1$, if we explore the regions of the geometry where $r = \Or[a]$,  we find a different answer. In coordinates where $r = a \xi$, the metric expanded to $\Or[\xi^2]$ near $\xi = 0$ is given by
\be
\label{xizerometric}
\begin{split}
ds_6^2 = &-\frac{\left(2 a^2-b^2\right) \left(\cos ^2(\theta )+\xi ^2\right)}{2 \lambda ^2} dt^2 +  \frac{b^2 \left(\cos ^2(\theta )+\xi ^2\right)}{ \lambda ^2} dt d y + \frac{\left(2 a^2+b^2\right) \left(\cos ^2(\theta )+\xi ^2\right)}{2 \lambda ^2} dy^2 \\ &+\lambda ^2 \left(1 - \xi ^2\right) d \xi^2 -2 \frac{a^2 \sin ^2(\theta )}{\sqrt{a^2+\frac{b^2}{2}}} dt d \phi - \frac{b^2 \cos ^2(\theta )}{ \sqrt{a^2+\frac{b^2}{2}}} dt d \psi - 2 \sqrt{a^2+\frac{b^2}{2}} \cos ^2(\theta ) d y d \psi \\
&+ \lambda^2 d \theta^2 + \lambda^2 \sin^2 \theta d \phi^2 + \lambda^2 \cos^2 \theta d \psi^2.
\end{split}
\ee

If we take $a \sim \lpl$, then the fuzzball metric has Planckian structures near $r = 0$ and these structures are given by the metric in \eqref{xizerometric}. Remarkably, curvature invariants such as the Ricci scalar and even the square of the Riemann tensor, $R_{\mu \nu \rho \sigma} R^{\mu \nu \rho \sigma}$, which can be computed from \eqref{xizerometric} remain {\em finite} in  limit as $a \rightarrow 0$.  So, in this limit, the fuzzball metric cleverly introduces Planckian structures, without introducing Planckian curvatures! This is a surprising and nice feature of the solution. 

However, it would be {\em incorrect} to imagine that this makes the classical metric immune to quantum fluctuations in this region. The study of section \ref{sectwocharge} tells us that we must also take into account quantum fluctuations in the {\em parameters} that specify a solution. In the two-charge case, the solution was specified by a profile function $F^i(s)$. But, in the quantum theory, the profile function did not have a definite value because of the non-zero commutator, $[F^i(s), F^i(\sp)]$.  We saw that when the metric had Planck scale features, the uncertainties in the profile function  were enough to make these features unreliable.

In the absence of a moduli-space quantization of the solutions examined in this section, we cannot make analogous precise statements for the multi-charge solutions.  However, a rule of thumb is that we do not expect to pin down bulk length-scales in a theory of quantum gravity with perfect certainty. Therefore, it is fair to estimate that in the quantum states corresponding to the metrics examined in this section, the length-scale $a$ will itself fluctuate and that $\delta a = \Or[\lpl]$.

If this is correct, then  in the regime where $a = \Or[\lpl]$ we also have ${\delta a \over a} = \Or[1]$. But then, examining the metric \eqref{xizerometric}, we see that such fluctuations will {\em induce} fluctuations in the metric so that $\delta g_{\mu \nu} = \Or[g_{\mu \nu}]$. For example, we see that for the determinant of the metric \eqref{xizerometric}
\be
g = \det(g_{\mu \nu}) = a^4 \lambda ^4 \xi ^2 \left(\xi ^4-1\right) \sin ^2(\theta ) \cos ^2(\theta ) ,
\ee
and therefore
\be
\delta g = {\partial g \over \partial a} \delta a = 4 g {\delta a \over a}.
\ee
If  ${\delta a \over a} = \Or[1]$ then ${\delta g \over g} = \Or[1]$. So, while the Planckian structures are smooth, in the sense that local curvature invariants remain bounded, they are nevertheless not reliable features of the metric.

We caution the reader that our arguments in this last paragraph have been necessarily somewhat imprecise. This is because the fuzzball program has itself not been carried through to completion in this setting. But we believe that our reasoning is robust for a simple reason.  The parameters that specify a fuzzball solution are coordinates on the phase space of gravity.  
Usually, we do not consider classical solutions whose distinctive features depend on specifying phase-space coordinates to an accuracy that depends on $\hbar$. Conversely, if we attempt to do so, we should expect that the minimal fluctuations on phase space induced by the uncertainty principle will wash out these features.

\section{Conclusions}

In this paper, we examined the fuzzball program by checking its consistency with some general results from statistical mechanics and some simple physical expectations. While the differences between typical black-hole microstates are exponentially suppressed, and so cannot be described geometrically, we showed that fuzzballs cannot even provide a reliable basis for the space of microstates.

We argued that fuzzballs that differ macroscopically from the conventional black hole are too atypical to be elements of the basis of microstates. We checked in section \ref{secmulticharge} that such macroscopic deviations can be detected by simple asymptotic correlators. We also showed that fuzzballs that cap off a macroscopic distance away from the horizon have a gap between the frequency of allowed excitations that is too large for black-hole microstates. We emphasize that although our calculations were limited to specific sets of geometries,  we expect this fact to generalize to arbitrary solutions that have macroscopic features.

These results imply that viable microstates must resemble the black hole very closely all the way up to Planck length from the horizon, where they must suddenly deviate from the black-hole geometry so that space ends before a horizon forms. However, we argued that in this region --- just  where fuzzballs start to show interesting deviations from the black-hole geometry  --- the solutions are also expected to become unreliable since quantum fluctuations  become of the same order as classical expectation values of components of the metric. 

We verified this expectation through a detailed consideration of the two-charge fuzzball geometries. Here, using the well known quantization of this set of solutions, we were able to compute the expectation values and the quantum fluctuations for the harmonic functions that enter the metric and for a one-form that probes the geometry. At the end of these involved calculations, we found that, just as expected, typical microstates start to deviate from the conventional geometry only at the Planck scale, and quantum fluctuations make the solutions unreliable precisely in this region.

In the two-charge case, the conventional geometry has no horizon. However, whenever conventional black holes do have a regular horizon, the black-hole geometry continues to be reliable well past the horizon of the black hole and so --- unlike the geometry of fuzzballs --- there is no reason to distrust the black-hole geometry at the horizon scale. So, our analysis supports the standard picture that the distinct microstates of a macroscopic black hole are all represented by the same geometry with a regular horizon.  

The fuzzball program is sometimes supported through indirect arguments --- by suggesting that the information paradox necessitates the existence of structure at the horizon. We explained in section \ref{secstatprelim} that these indirect arguments rely on an assumption of exact locality in quantum gravity. But extensive evidence from AdS/CFT and string theory suggests that this assumption is incorrect.

Our results are consistent with the results of Sen \cite{Sen:2009bm}. Sen pointed out that, in the context of the two-charge system, in the Type II-B duality frame, the ground  states of the D1-D5 system could not be represented by a string-scale black hole. This led Sen to suggest  that fuzzballs should be understood as parameterizing the ``hair'' around a black hole, and not the microstates of the black hole itself. Where fuzzball solutions exist, Sen's argument suggests that quantizing the moduli-space of fuzzballs should not be expected to reproduce the Bekenstein-Hawking entropy; rather the entropy obtained by quantizing classical supergravity solutions should be {\em added} to the Bekenstein-Hawking entropy to obtain the full degeneracy of the system.

Our results rely on some physical assumptions, two of which are outlined in expectations \ref{continuousspectrum} and  \ref{euclideanaverage}. For the fuzzball program to be correct, some of these assumptions would have to be violated. For example, if it could be shown that, for some reason, even the Euclidean black hole is an inadmissible saddle-point to compute thermal averages, then this could be used to invalidate our estimate of fluctuations in the metric in equation \eqref{smallsigmaens}. This would open the door to allowing fuzzballs that can be described classically, and perhaps differ from the conventional solution at the string scale.  

At the same time, the fuzzball program would also have to explain why two-point functions of asymptotic operators computed in fuzzball microstates do not display a continuous spectrum. In AdS/CFT, this would  suggest that either the boundary theory has a very exotic set of couplings  --- so that fuzzball microstates span the microcanonical ensemble but yet are not connected to most nearby energy eigenstates by the action of light primary operators  --- or that exact degeneracies in energy eigenvalues are somehow restored at the supergravity point. These possibilities seem rather implausible.

It seems more plausible to us that  fuzzballs should simply be thought of as ``stars'' with the same charges as black holes and not as microstates of the black hole. In this context, fuzzballs  are very interesting solutions. The idea of stabilizing a self-gravitating system against gravitational collapse by causing space to end before a horizon is formed is also remarkably rich. The asymptotically AdS fuzzball solutions that have been found so far appear to represent valid states in the AdS/CFT correspondence, and it is an interesting problem to study their properties and determine the CFT duals to these states.

\section*{Acknowledgments}
We would like to thank Junggi Yoon for collaboration in the early stages of this work. We are grateful to Iosif Bena, Jan de Boer, Emil Martinec, Bidisha Chakrabarty, Monica Guica, Stefano Giusto,  Dileep Jatkar, Jared Kaplan,  Samir Mathur, Shiraz Minwalla, Yasunori Nomura, Kyriakos Papadodimas, Andrea Puhm,  Ashoke Sen, Rodolfo Russo, Yogesh Srivastava, Kostas Skenderis, Bo Sundborg, Marika Taylor, Amitabh Virmani and Nick Warner for helpful discussions. We are grateful to Iosif Bena and Masaki Shigemori for sharing their notes on the solutions of section \ref{secmulticharge}.  We are grateful to Monica Guica and Ashoke Sen for comments on a draft of this paper. S.R. is partially supported by the Swarnajayanti fellowship of the Department of Science and Technology. S.R. would like to acknowledge the hospitality of the Kavli Institute for the Physics and Mathematics of the Universe (Tokyo), the Tata Institute of Fundamental Research (Mumbai), the National Institute for Science Education and Research (Bhubaneshwar), and the Institute of Theoretical Physics (Saclay)  for hospitality while this work was being completed. We would also to acknowledge the Kavli Asian Winter School, ICTS/Prog-KAWS2018/01, for hospitality and thank all the participants of the Bangalore Area Strings Meeting,   ICTS/Prog-basm2017/2017/07 for discussions.

\appendix

\bibliographystyle{JHEP}
\bibliography{references}
\end{document}